\documentclass[useAMS,usenatbib]{mn2e}
\usepackage{graphicx}
\usepackage{epsfig}

\newcommand{\HI}{H\,{\sc i}}
\newcommand{\HII}{H\,{\sc ii}}
\newcommand{\SII}{[S\,{\sc ii}]}

\newcommand{\OIII}{[O\,{\sc iii}]}
\newcommand{\OIV}{[O\,{\sc iv}]}
\newcommand{\OII}{[O\,{\sc ii}]}
\newcommand{\OI}{[O\,{\sc i}]}

\newcommand{\NII}{[N\,{\sc ii}]}
\newcommand{\HeII}{He\,{\sc ii}}
\newcommand{\HeI}{He\,{\sc i}}

\newcommand{\FeII}{[Fe\,{\sc II}]}
\def\p0{\phantom{0}}

\title[New PNe in the Large Magellanic Cloud (IV)]{A new population of planetary nebulae discovered in the Large
Magellanic Cloud (IV): the outer LMC}
\author[Warren A. Reid and Quentin A. Parker]{Warren A. Reid$^{1,}$$^{2}$\thanks{e-Mail:
warren.reid@mq.edu.au; war@aao.gov.au } and Quentin A.
Parker $^{1,}$$^{2,}$$^{3}$\thanks{e-Mail: quentin.parker@mq.edu.au}\\
$^{1}$Department of Physics and Astronomy, Macquarie University, Sydney, NSW 2109, Australia\\
$^{2}$Centre for Astronomy, Astrophysics and Astrophotonics, Macquarie University, Sydney, NSW 2109, Australia\\
$^{3}$Anglo-Australian Observatory, PO Box 296, Epping, NSW 1710
Australia}
\begin{document}

\date{}

\pagerange{\pageref{firstpage}--\pageref{lastpage}} \pubyear{2002}

\maketitle

\label{firstpage}

\begin{abstract}
We have extended our PNe survey to the outer $\sim$64deg$^{2}$ of the Large Magellanic Cloud (LMC) using maps from the Magellanic Cloud Emission Line Survey (MCELS) and the UK Schmidt Telescope (UKST) H$\alpha$ survey. Although the MCELS survey has poorer $\sim$5~arcsec resolution than both the UKST H$\alpha$ survey and the original H$\alpha$ median stacked map in the LMC's central 25deg$^{2}$, it has the advantage of additional narrow-band filters at H$\alpha$, \OIII~and \SII~providing improved diagnostic capabilities. Using these data to uncover new emission line candidates we have so far spectroscopically confirmed an extra 61 LMC PNe which we present here for the first time. We have also independently recovered and spectroscopically confirmed 107 of the 109 (98\%) PNe that were previously known to exist in the outer LMC. The majority of our newly discovered and previously known PNe were confirmed using the AAOmega, multi-object fibre spectroscopy system on the 3.9-m Anglo-Australian Telescope (AAT) and the 6dF multi-object spectrograph on the UKST. These newly identified PNe were cross-checked against extant multi-wavelength imaging surveys in the near and mid-infrared in particular and against the latest emission-line ratio diagnostic plots for improved confidence in PNe identification.

\end{abstract}


\section{Introduction}
The establishment of the most complete sample of spectroscopically confirmed PNe in the LMC is of enormous value for population studies, luminosity functions, chemical abundance and evolutionary studies. Working on such a sample of LMC PNe and across the widest physical extent of this galaxy also gives us a unique insight into the kinematic motions of the older LMC populations which can be compared to young populations and the \HI~disk of the LMC (eg. Reid \& Parker, 2006b). With the LMCs well established 50Kpc distance (Reid \& Parker 2010 and references therein) low and relatively consistent reddening of about E(V-I) = 0.09 $\pm$ 0.07 mag using the red clump method (eg. Haschker et al. 2011), modest 35 degree angle of inclination and $\sim$500pc disk thickness (van der Marel \& Cioni 2001), LMC PNe have many advantages over similar studies using Galactic PNe.

Our UKST H$\alpha$ deep-stack survey of the central 25deg$^{2}$ of the Large Magellanic Cloud (Reid \& Parker 2006a) was initially used to identify faint emission-line candidates of all kinds for spectroscopic confirmation. From this survey and the subsequent spectroscopy we previously identified 462 PN candidates and confirmed 170 previously known PNe in the survey area (Reid \& Parker, 2006a,b). We have also published our first catalogue of newly identified LMC emission line stars (Reid \& Parker, 2012). We later modified the PN numbers to 411 new and 164 previously known PNe with the aid of further spectroscopic follow-up and the commencement of multi-wavelength and emission-line ratio comparisons (Reid \& Parker 2010). Such comparisons offer additional diagnostic capability to weed out contaminating PN mimics as shown by Frew \& Parker (2010) and Sabin et al. (2011).

We have now extended our survey to cover 40deg$^{2}$ of the outer LMC beyond our original central 25deg$^{2}$ field using the 64deg$^{2}$ coverage given by the Magellanic Cloud Emission-line Survey (MCELS)~\footnote{C. Smith, S. Points, the MCELS Team and NOAO/AURA/NSF}. In addition we have access to the higher resolution single exposure Anglo-Australian Observatory/UK Schmidt Telescope (AAO/UKST) H$\alpha$ and short red band `SR' survey fields that completely overlap with the extended MCELS region. These wide-field images were exposed between 1997 and 2001 using tech-pan film on the UKST for eventual inclusion in the on-line SuperCOSMOS R and H$\alpha$ sky surveys. The digitised pixel size is 10 micron (0.67 arcsec) and is provided in density units, scaled to maximize the range of 16-bit integers per pixel provided the SuperCOMSOS. This additional digital data covering the Magellanic Clouds and Magellanic stream is not currently available on line but our access to it has enabled us to verify, at higher angular resolution, many of the new candidate MCELS objects as genuine emission sources. These candidates were followed up spectroscopically using AAOmega on the AAT, 6dF on the UKST and the 1.9m telescope at the South African Astronomical Observatory (SAAO). The resulting reduced spectra were then measured and examined to determine the strength of key emission lines and ratios such as \NII/H$\alpha$, \OIII5007\AA/H$\beta$ and \HeII4686\AA/H$\beta$ as described in Reid \& Parker (2006a,b).

Once strong candidates were found, they were examined using false colour images from the $\textit{Spitzer}$ SAGE survey (Surveying the Agents of a Galaxy's Evolution) (Meixner et al. 2006) and 2MASS (Skrutskie et al. 2006) (for bright candidates) in order to determine the infrared excess. The combination of these spectroscopic criteria and multi-wavelength images have allowed us to confidently assess and classify PN candidates as necessary. Additional candidate source photometry was also gained from the Magellanic Cloud Photometric Survey (MCPS) (Zaritsky et al. 2004) depending on coverage and magnitude limits. A combination of all of the above photometric and spectroscopic data has allowed us to decisively extend our PN survey to the sparser outer regions of the LMC and provide firmer probability ratings as to the true nature of the emission sources presented in this paper. Our previous LMC PNe samples reported in Reid \& Parker (2006a,b) have also been re-evaluated in the light of these new diagnostic capabilities. The results of this investigation and the accompanying study into the multi-wavelength characteristics of the entire known LMC PNe sample is presented in the sister paper to this (Reid \& Parker, 2013). This additional study has resulted in the re-classification or strengthened confirmation of many of the less secure PNe identifications (originally published as `PN possible' or `PN likely') from our previous catalogue.

In section~\ref{section2} we briefly describe the MCELS survey, the AAO/UKST LMC H$\alpha$ survey and our method of candidate selection. In section~\ref{section3} we describe the 2dF and 6dF confirmatory followup observations including data reduction, and details regarding the de-reddening of the fibre spectra.  Section~\ref{section4} describes the diagnostic methods employed to evaluate spectral emission lines, our results for extinction, multi-wavelength checks undertaken. We then present diagnostic comparison plots using optical emission lines and provide the table of the newly discovered  LMC PNe. Section~\ref{section5} gives a summary and a brief description of our continuing object followup of newly discovered emission candidates and some conclusions.


\section{The candidate selection technique}
\label{section2}
\subsection{The MCELS survey}
The MCELS survey was designed to identify and study the interstellar medium (ISM) within the Magellanic Clouds as well as discrete emission objects such as \HII~regions, supernova remnants, planetary nebulae, superbubbles and giant shells. It was intended that it be used as a foundation and applied together with other multi-wavelength surveys and observations (Smith et al. 1998).

The survey employed large-format CCD detectors on the Schmidt telescope at CTIO to create uniform data sets which imaged the central 8 $\times$ 8deg$^{2}$ of the LMC and the central 3.5deg $\times$ 4.5deg$^{2}$ of the SMC. This produced 3 arcsec/pixel data with a resolution of $\sim$5 arcsec. Individual exposures covered a 1.1deg $\times$ 1.1 field of view. Areas were sampled an average of eight times in four different regions of the detector to assist in eliminating flat fielding and other instrumental effects.

Exposures were taken using custom-made 4-inch interference filters centred on H$\alpha$~($\lambda$6563, $\Delta\lambda$ = 30\,\AA), \SII~($\lambda$6724, $\Delta\lambda$ = 50\,\AA) and \OIII~($\lambda$5007, $\Delta\lambda$ = 40\,\AA). Continuum-band filters at $\lambda$6850, $\Delta\lambda$ = 95\,\AA~and $\lambda$5130, $\Delta\lambda$ = 155\,\AA were also obtained to allow subtraction of stellar sources.

\subsection{The AAO/UKST H$\alpha$ survey}
The AAO/UKST H$\alpha$ survey (1998-2003) used the world's largest monolithic narrow-band H$\alpha$ interference filter (Parker \& Bland-Hawthorn 1998) with a 70\AA~full width at half-maximum (FWHM) bandpass (so actually a H$\alpha$ plus \NII\,6548, 6584\AA~filter). It was the last, great UKST survey, the only one to be taken on Kodak Technical Pan film rather than glass plates (Parker \& Malin 1999) and the only one to be available to the community solely in digital form\footnote[1]{http://www-wfau.roe.ac.uk/sss/halpha/}. The main survey covered 4500deg$^{2}$ of the Southern Galactic plane (Parker et al. 2005) and has led to a wealth of PN discoveries in its own right (e.g. Parker et al. 2006, Miszalski et al. 2008). However, at the same time, a little known H$\alpha$ survey of 40 UKST fields in and around the Magellanic Clouds (including the bridge and stream) was also undertaken by Parker \& Morgan. This was in addition to their multi-exposure (12 H$\alpha$ and six SR) deep stack over the central UKST field of the LMC as already reported (Reid \& Parker 2006a,b).

We have special access to the 40 field AAO/UKST LMC and SMC H$\alpha$ survey which includes matching short red exposures. In this work it has been used extensively to confirm emission source candidates uncovered by MCELS. However, it can also be used for emission line source candidate selection outside the 64deg$^{2}$ area covered by the MCELS survey and as such will form the basis of a future investigation by our team. The SuperCOSMOS LMC H$\alpha$ map has proved particularly useful in cases where bad pixels in the MCELS maps often resembled point emission sources. The 5 Rayleigh sensitivity and arcsec spatial resolution of the UKST based Magellanic cloud H$\alpha$ survey meant that these data were able to provide some vital morphological information to assist our source selection criteria and spectroscopic target prioritization.

\subsection{PNe candidate selection}
\begin{figure}
\begin{center}
  \includegraphics[width=0.48\textwidth]{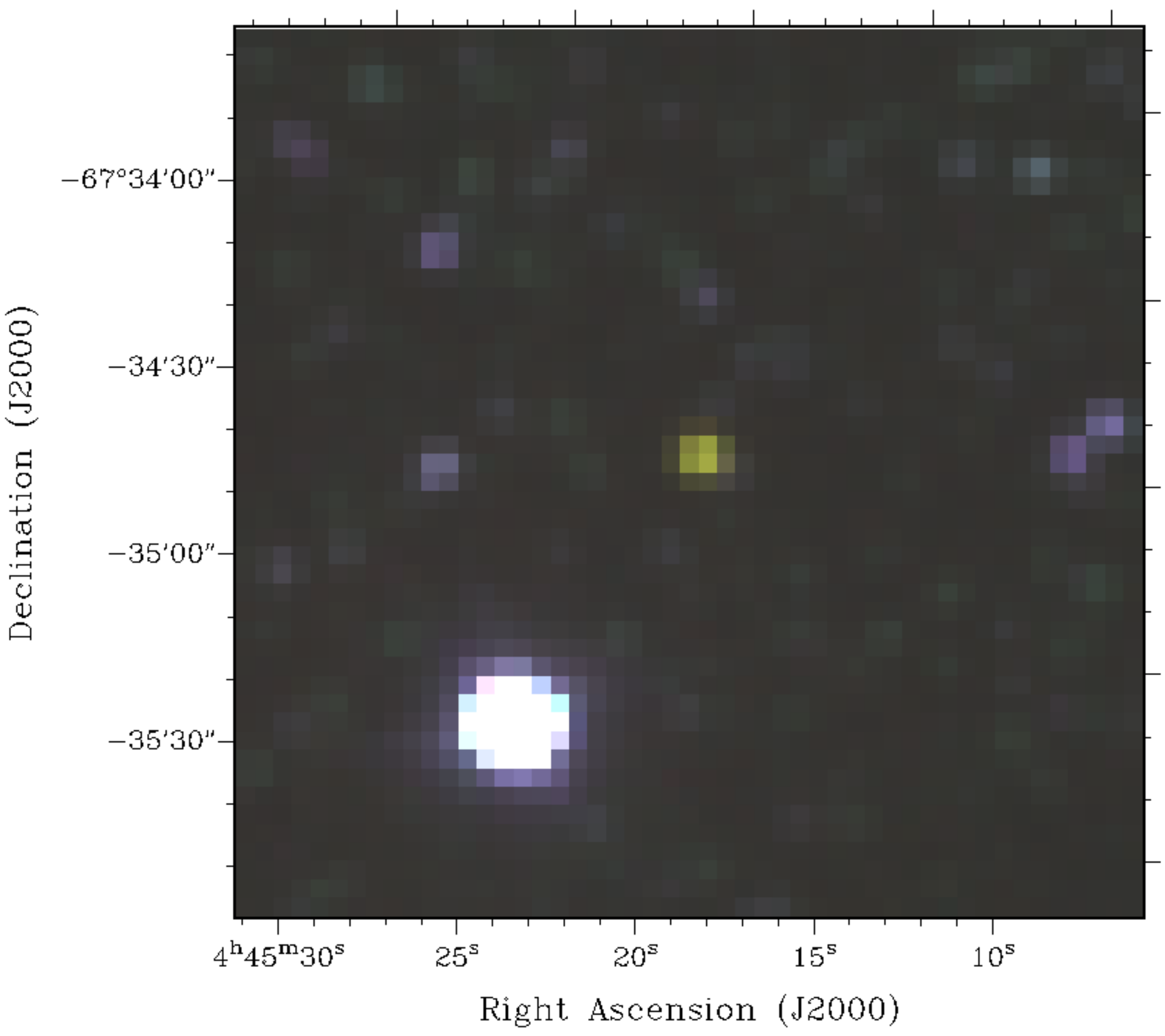}\\
  \end{center}
\begin{center}
  \includegraphics[width=0.48\textwidth]{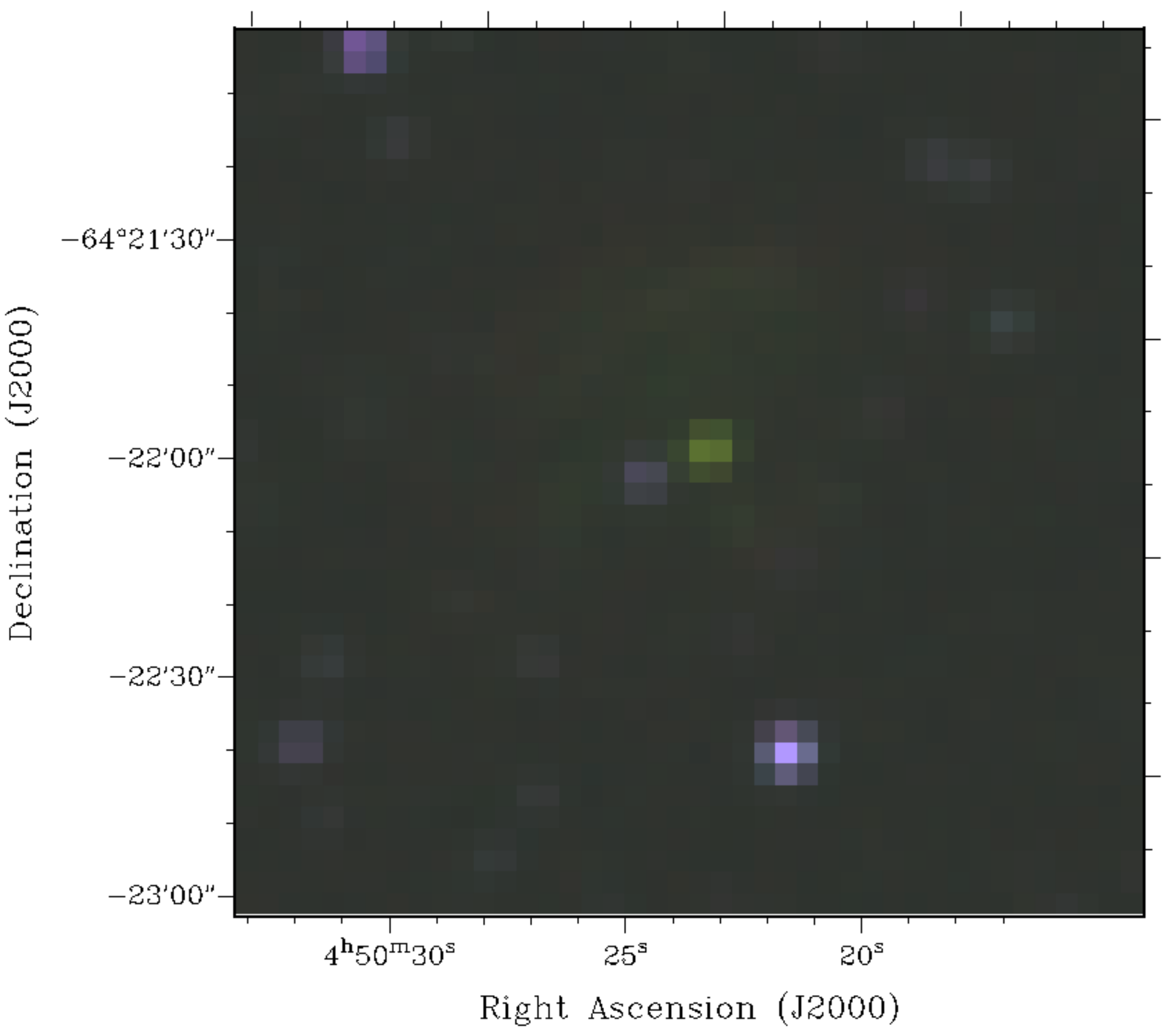}\\
  \end{center}
  \caption{The combined H$\alpha$ (red), \OIII5007\AA~(green) and \SII6724\AA~(blue) MCELS image of previously known PNe Mo5 above and Mo8 below. The various yellow-green-brown, and even pinkish colours of the PNe, produced by the particular combination of emission lines that contribute to the detected signal in each filter, is typical of previously identified PNe.}
  \label{Figure1}
  \end{figure}

\begin{figure}
\begin{center}
  \includegraphics[width=0.48\textwidth]{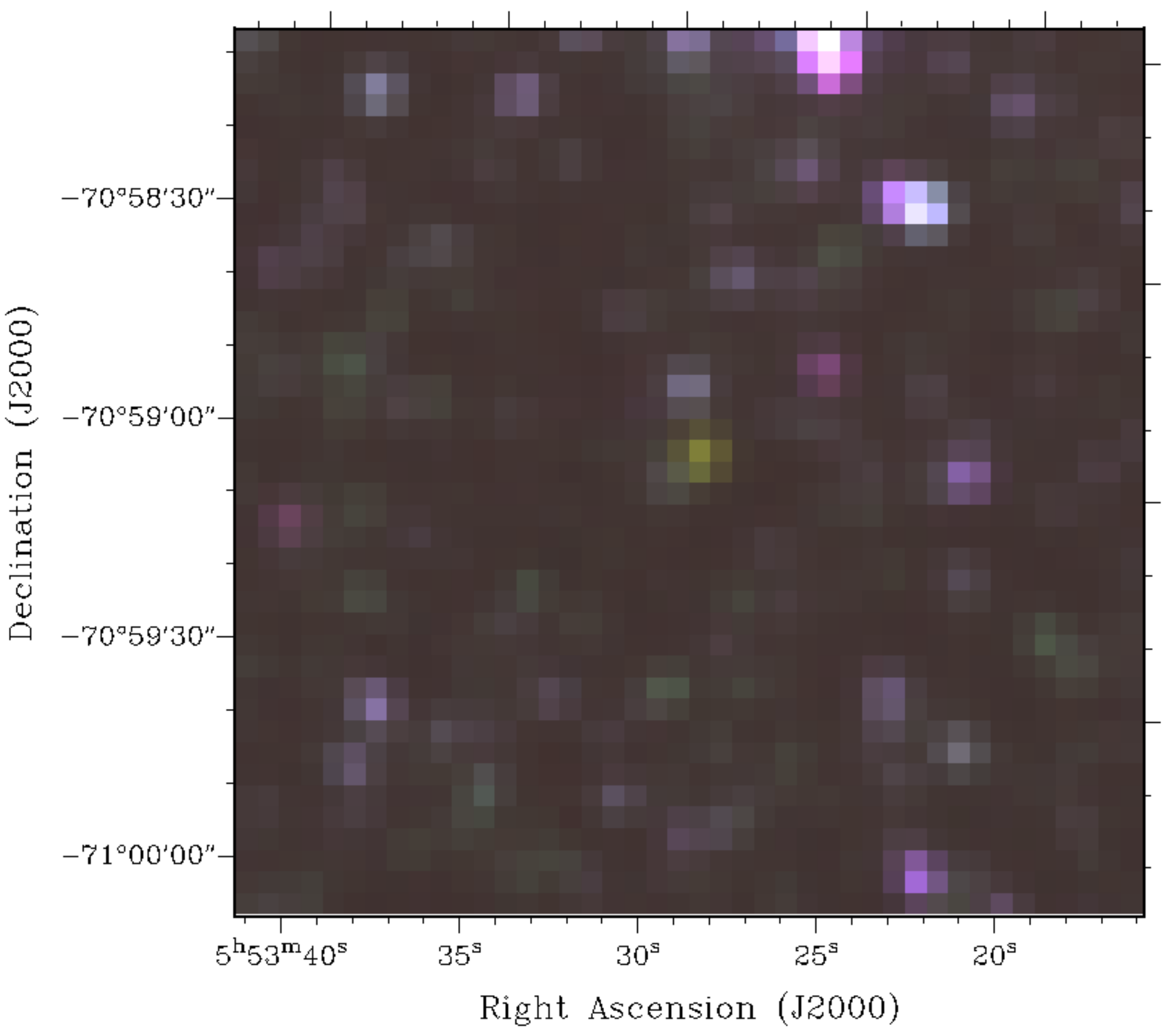}\\
 \end{center}
\begin{center}
  \includegraphics[width=0.48\textwidth]{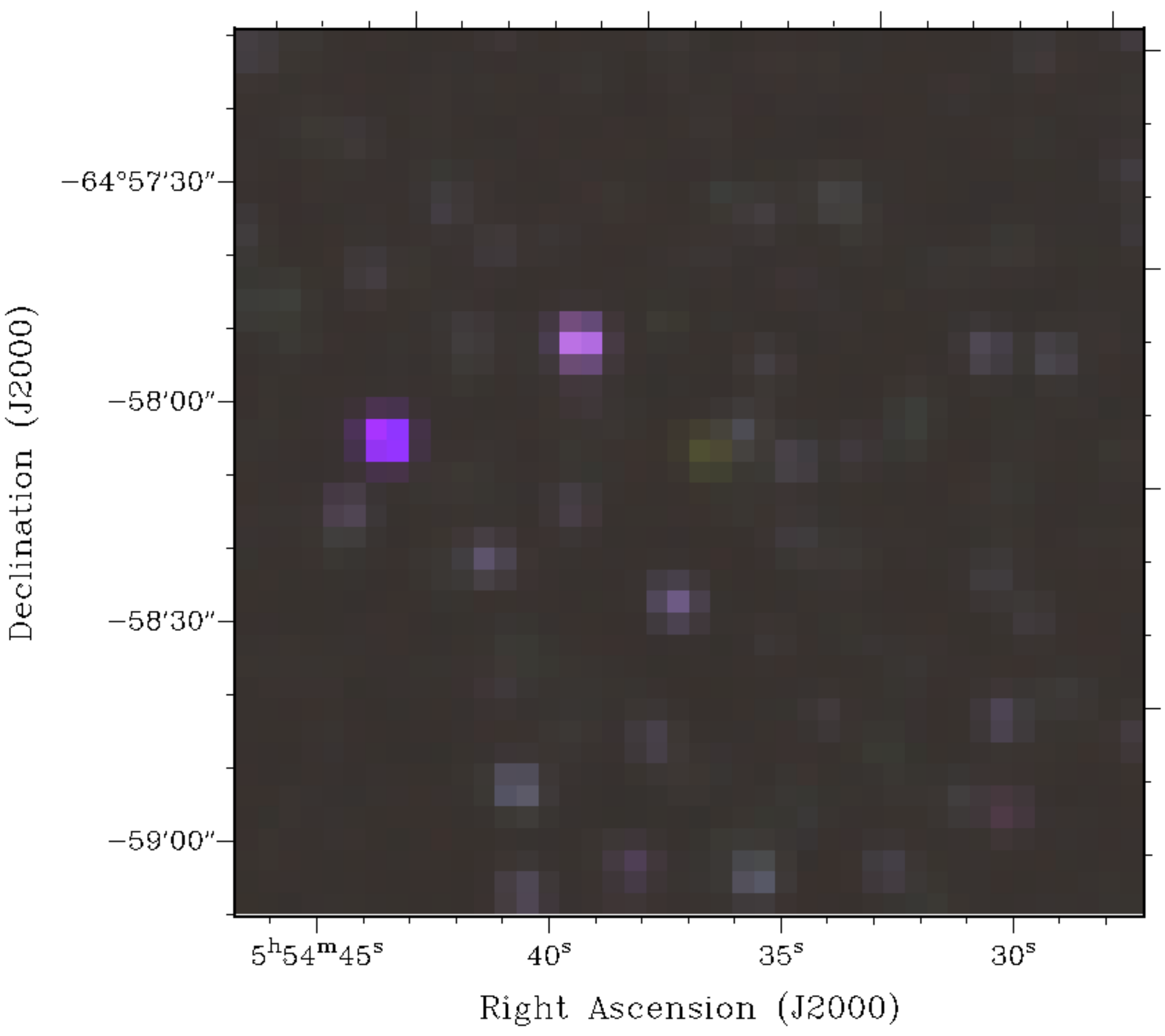}\\
  \caption{The combined H$\alpha$ (red), \OIII5007\AA~(green) and \SII6724\AA~(blue) MCELS image of a selection of confirmed new PNe RP4151 and RP4176. The faint yellow-brown colour of the PNe, produced by the particular combination of emission lines that contribute to the detected signal in each filter, is used to identify new candidates.}
  \label{Figure2}
  \end{center}
  \end{figure}

Candidate emission sources were selected from the MCELS maps by assigning a false RGB colour to each of the three narrow-band filters and merging them into one colour map. We then searched the combined colour map for sources that indicated high \OIII$\lambda$5007${\mbox\AA}$ and H$\alpha$ signatures. Candidate PN sources were those that (in terms of their colours) fell within these two bands and which are expected to comply with expectations for LMC PNe in low extinction, low metallicity environments. For example, by assigning red to H$\alpha$, green to \OIII~and blue to \SII, all bright PNe appear as yellow-brown/green sources since yellow is a mixture of red and green. Normal, continuum sources (stars) appear with colours varying between blue, white and red, providing a clear differentiation between them and potential emission objects. Examples of previously known and newly identified LMC PN uncovered in this way are given in Figs~\ref{Figure1} and ~\ref{Figure2} respectively. This MCELS PN candidate recognition process was significantly aided by assessing the new candidate PNe with respect to the appearance of known PNe using the same merged colour identification technique. The identification and initial assessment was supplemented by our special access to SuperCOSMOS multi-exposure H$\alpha$/SR exposure images covering the regions outside the original 25deg$^{2}$ LMC region.

This somewhat manual method of object detection using both MCELS and SuperCOSMOS merged colour imaging proved to be more effective, allowing better discrimination than the method provided by the standard {\footnotesize DAOPHOT} and {\footnotesize SeXtractor} software packages. These packages are especially good at finding faint galaxies through a process called segmentation, which identifies regions which differ in brightness, colour, texture or can be delineated by edges. In short, these packages use two images to look for a combination of fluxes which are different to other objects. Since PNe can be very bright or faint in the \OIII5007\AA~and the (H$\alpha$ + \NII)~bands with a variety of ratios between the two and widely differing strengths of \SII, there is an extremely wide variation in possible colour combinations. With the low resolution of the MCLES maps where most PNe occupied only 2 - 4 pixels, it proved more efficient to train the eye to detect both bright and faint candidates dominated by either \OIII~or \NII. This allowed us to include all candidates to the faintest detectable limits of the survey (\OIII= 1.4$\times$ 10$^{-17}$ erg cm$^{-2}$ s$^{-1}$ arcsec$^{-2}$; H$\alpha$= 7.0$\times$ 10$^{-18}$ erg cm$^{-2}$ s$^{-1}$ arcsec$^{-2}$; \SII= 5.2$\times$ 10$^{-18}$ erg cm$^{-2}$ s$^{-1}$ arcsec$^{-2}$, in agreement with Pellegrini et al. (2012)), while giving a low priority to probable emission-line stars and compact \HII~regions. Nonetheless, the central positions for all emission objects uncovered from the maps were carefully recorded for future spectroscopic and multi-wavelength followup.

Very faint PNe remain difficult to detect in the MCELS data. This is because the survey does not go as deep for point source detection in the three bands as our central H$\alpha$ arcsecond resolution median-stacked data and because MCELS has poorer angular resolution ($\sim$5\,arcsec), with pixels as large as 3~arcsec on the sky. This means that, in many cases, faint PNe in the LMC MCELS data are not much larger than 1 pixel across. Although the survey aimed to reach a limiting surface brightness of approximately 3 $\times$ 10$^{-17}$ erg cm$^{-2}$ s$^{-1}$ arcsec$^{-2}$ (Smith et al. 1999), the low resolution means that any point sources near that level are impossible to define. Detection is also significantly degraded towards the edges of the MCELS field where bad pixels interfere with real detections.

\begin{figure}
\begin{center}
  \includegraphics[width=0.5\textwidth]{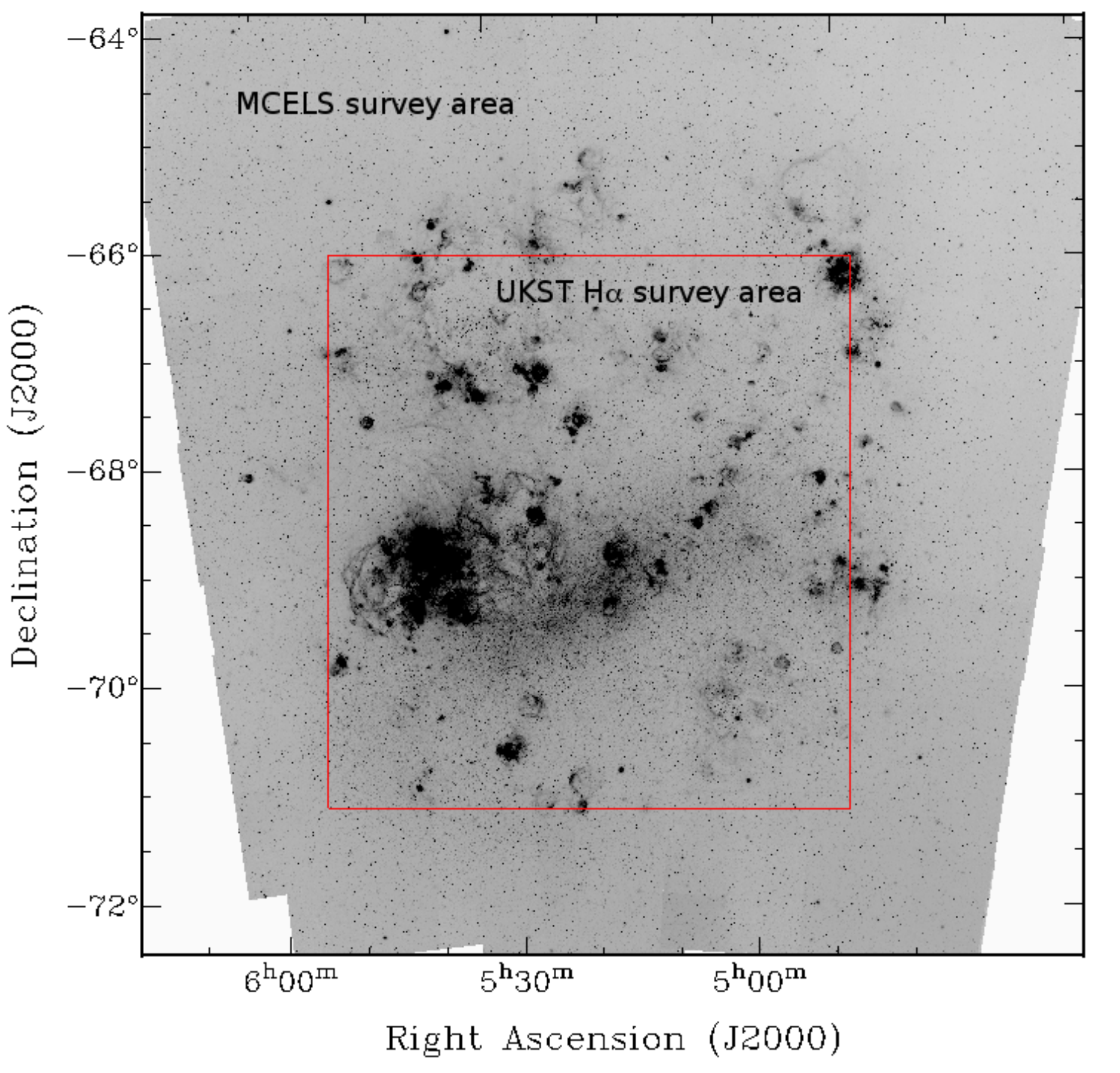}\\
  \caption{A comparison of the areal LMC coverage provided by the MCELS survey (larger area) and the UKST H$\alpha$ 25deg$^{2}$ deep-stack survey area (bounded by the red box).}
  \label{Figure3}
  \end{center}
  \end{figure}

Despite these drawbacks, careful scrutiny of the combined colour images for sources with the tell-tale PNe-like colours provided a large number of targets for spectroscopic followup. Almost half of these new MCELS candidates showed a central (possibly stellar) source, while the rest gave the impression of an extended nebulae. About 90\% of the targets were drawn from the outer LMC which was not covered in the original deep stack UKST H$\alpha$ survey of the central LMC (Reid \& Parker, 2006a,b). The others were objects that were subsequently found using both MCELS and our original UKST H$\alpha$ deep stack. The additional area covered by MCELS compared to our original survey region can clearly be seen in Fig~\ref{Figure3}~which uses a red line to indicate the maximum extent of our original UKST H$\alpha$ survey deep median stack.

Once a discrete source was found in the MCELS data, it was also examined using other available multi-wavelength data such as the broad-band optical SuperCOSMOS Sky Survey (SSS) $\textsl{B-R-I}$ images, 2MASS in the Near-IR (NIR), SAGE and $\textsl{WISE}$ in the mid-IR (MIR). These data allowed us to not only confirm the emission sources but check relative band strengths, where coloured and merged bands often result in specific combinations which are typical of PNe whose multi-wavelength characteristics and emission lines betray their presence in these passbands (e.g. see Parker et al. 2012).

As previously mentioned we also now have access to the 40 single-exposure UKST H$\alpha$ fields in and around the Magellanic clouds. Careful examination of one of these fields in the far outer regions of the LMC and beyond the extent of the MCELS survey has uncovered a few additional candidate sources. These survey fields, which are not yet generally released nor in the public domain, were each assigned a distinguishing colour and merged to reveal the H$\alpha$ excess in exactly the same way as originally reported in Reid \& Parker (2006a) for the UKST deep median stack data. Furthermore, the higher resolution of $\sim$1-1.5~arcsec of the UKST H$\alpha$ data (with 0.67~arcsecond SuperCOSMOS pixels) provides improved resolution, crucial for indicating the presence of any faint extensions such as jets and possible AGB haloes as well as very faint, compact yet resolved nebulae. This small pilot study indicates there are many emission line sources yet to be detected in the extensive Magellanic Cloud UKST
H$\alpha$ survey region.

\section{AAT AAOmega and UKST 6dF spectroscopic observations}
\label{section3}

Details of the spectroscopic follow-up of the newly uncovered MCELS emission-line candidates as well as a description of the data reduction technique and corrections for extinction and reddening are given below.

\subsection{AAOmega observations}
A four night observing run on the Anglo-Australian Telescope (AAT) from 2010 February 9 to 12 was undertaken using the 2dF multi-object fibre positioner (Lewis et al. 2002) coupled with the AAOmega spectrograph (Saunders et al. 2004; Sharp et al. 2006).  These data were supplemented by AAOmega service time observations of single 2dF fields on the 2012 February 14, 15 and 17.

The powerful 2dF system was an ideal instrument for the spectroscopic followup of large numbers of
MCELS candidate emission-line objects due to its  ability to simultaneously observe
400 targets (including objects, fiducial stars and sky positions) with 2~arcsec fibres over the wide 2$^{\circ}$ diameter field area, enabled by the AATs top-end corrector. This large six-element transmissive corrector lens-system incorporates an atmospheric dispersion compensator, which is essential for wide wavelength coverage using small diameter fibres couple with long exposure times.

The combined observations provided a total $\sim$2,693 spectra. The large number of available fibres allowed even dubious PNe candidates to be observed while also allowing significant new samples of \HII~regions and emission line star candidates to be targeted.  As a result, we were able to confirm 61 sources as new LMC PNe.

Each AAOmega field was observed twice, each with an exposure time of 1200s. The resulting 1200s x 2 exposures for each field setup allowed us to combine exposures and remove cosmic rays. AAOmega was used to obtain simultaneous red and blue arm observations with the 580V and 1000R gratings split using a dichroic. Central wavelengths of 4800\AA~and 6800\AA~were chosen to cover the most important diagnostic lines in each arm.
For the blue arm the wavelength range was 3736-5854\AA~at a dispersion of 4.30\AA/pixel and for red arm the wavelength range was 6213-7373\AA~at a dispersion of 0.552\AA/pixel. These medium-resolution observations, at 1.023\AA~FWHM for V and 1.905\AA~FWHM for R, were used as the primary means of object classification and provided the bright \OIII5007\AA~and H$\alpha$ fluxes for this study. These resolutions were sufficient for kinematic studies and were also used to measure the \SII~6716,6731\AA~lines for determination of nebula electron densities.

\subsection{6dF observations}
Two nights of dedicated 6dF observations on the UKST were undertaken by the first author on 2012 February 3 and 4 (eight fields in all). These multi-fibre observations using the standard 580V and 425R gratings with a dispersion of 0.62 \AA/pixel, provided spectroscopic confirmation of 562 additional sources while overlapping with AAOmega fields. The observations resulted in the confirmation of 38 new LMC PNe, five of which were only observed solely through 6dF observations. Eight overlapping 6dF fields were observed, each of which allowed us to place $\sim$100 $\times$ 6 arcsec diameter fibres across a 5$^{\circ}$.7 field of view. An adaptive tiling algorithm and simulated annealing for field configuration were employed to ensure that the maximum number of fibres was allocated to targets and dedicated sky positions. The employed $\textsl{V}$ and $\textsl{R}$ gratings were swapped in turn without any need to change the field plate or the tracking of the telescope. Two 1500s exposures were made on each field and with each grating setup. Both gratings provided a resolution of $\sim$1000 over the 3900--5590\AA~range for $\textsl{V}$ and the 5300--7570\AA~range for $\textsl{R}$ with a general signal-to-noise ratio (S/N) $\sim$10 per pixel achieved for continuum sources.


\subsection{Data reduction}
The 2dF and 6dF raw data were processed using the separate but related AAO 2dF/6dF data reduction pipelines 2dFDR\footnote[1]{http://www.aao.gov.au/AAO/2df/software.html\#2dfdr} and 6dFDR. These software pipelines can automatically reduce 2dF and 6dF multi-spectral frames creating the necessary calibration groups (eg. BIAS, DARK, FLAT, ARC, etc.). As a calibration exposure is reduced it is inserted into
the appropriate group. Before commencing the automated reduction process it is necessary to pre-select the relevant program files dedicated to each telescope, instrument and grating setup. Manual intervention and interaction is possible at various stages of the reduction process (e.g. Sharp et al. 2006).

For our data the software was instructed to perform a subtraction of background scattered light prior to the extraction.
The background was determined by fitting a function through the `dead' or unused fibres in the
image. Bias frames were obtained for each observed field. The bias strip was removed from the bias frame, trimmed and subtracted from the data.

The data reduction system performed a wavelength calibration using
the information from the spectrograph optical model. This was then
refined using data from the arc lamp exposure. The lines in the arc
lamp exposure were matched against a line list. A cubic fit for each
fibre was then performed to the predicted and measured wavelengths of
all lines which are non-blended, not too wide and not too weak. This
fit was then used to further refine the wavelengths.

To perform the sky subtraction, the data were first corrected for the relative fibre throughput, based on a throughput map derived from
the dedicated sky fibres.  The relative intensities of the skylines in the object data frame were used to determine the relative fibre
throughput. This method saved time, as no offset-sky observations were required.

The median sky (continuum subtracted) spectrum was calculated from the median of all the individual fibre sky spectra taken. For 2dF a minimum of 20 well-spaced sky fibres are usually chosen and verified to be true sky regions. This is particularly important for crowded field like the LMC as sky fibres contaminated by real sources (stars or nebulae) will affect the validity of the sky-subtraction.

Sky fibre spectra are normalized by their mean flux. A robust least-squares fit of the counts was
performed in the differential or continuum subtracted data fibres
versus the counts in the differential or continuum subtracted median
sky. Assuming that the sky was the dominant source of emission, the
slope of this fit gave the fibre throughput. The robust fit is
especially important when dealing with faint objects. The sky fibre
spectra in the data were then combined and subtracted from each
fibre.

It is normal for cosmic rays to be rejected automatically during the reduction process as each 2dF field observed usually has a minimum of two exposures which permit efficient rejection routines to be employed.

\subsection{Flux calibration and de-reddening of the 2dF and 6dF fibre spectra}

Since the PNe were observed with AAOmega and 6dF, a reliable flux calibration of the LMC PN emission-lines was required in order to properly compare optical spectra from different fibre-based observations. This also allows us to make meaningful comparisons between fibre spectroscopy and long-slit observations of individual objects, and to create diagnostic plots. In our previous work we reported on a new technique to obtain accurately calibrated spectral fluxes from the separate blue and red 2dF and 6dF exposures (Reid \& Parker, 2006a,b).

In the case of these new AAOmega observations a slightly different flux calibration process was required due to the changed instrument configuration compared to our earlier data. To calibrate the observed data counts for the same source observed simultaneously in the blue and red arms of the spectrograph, we used PNe with extremely low or non-existent continuum levels and well-determined emission-line fluxes gained from $Hubble~Space~Telescope$ (HST) observations (see Reid \& Parker 2006a, 2006b, 2010). These objects were deliberately included and observed on each field plate for use as flux calibrators for each individual field.

The process involved matching each spectral line on each field plate from each {\scriptsize CCD} camera
to raw PN fluxes gained from $HST$ exposures. Fluxes for LMC PNe from other
catalogues (Jacoby et al. 1990, Leisy et al. private communication,
Meatheringham et al. 1988) were also included in order to build up
the number of calibrators per field and improve error estimates. In order to extend the reliability of the calibration to faint lines of magnitudes $>$19, the \OIII4959\AA, H$\beta$, \HeII4686\AA,~\NII6584,6548\AA~and \SII6716,6731\AA~lines were included in the calibration. The individual AAOmega and 6dF line intensities for known PNe observed on each {\scriptsize CCD} and each field plate exposure were
 plotted against $HST$ and other published fluxes for the same lines (see Fig. 2 in Reid \& Parker, 2010). In each case, a line of best fit was derived and the underlying linear equation extracted.

 This equation became the calibrator for each emission line in each object where the {\scriptsize CCD} and individual 2dF field
 plate exposure were the same. The agreement of flux-calibrated PNe from each spectrograph/field plate combination was considered robust enough (within 0.15 dex) to allow calibration to all the blue and red emission lines for other emission objects observed in the same field. Full details including a discussion on the reliability of the method and error estimates are presented in Reid \& Parker (2010).

  \subsection{Corrections for extinction and reddening}

Extinction of light from distant planetary nebulae is the result of
dust both in the ISM and surrounding the PN itself. In each case, the light is both scattered and absorbed, which
increases the interstellar extinction towards shorter wavelengths.
The amount of extinction will differ for each object and needs to be
corrected to gain true fluxes. For our optical line fluxes, the
H$\alpha$/H$\beta$ ratio was used to determine the extinction
constant $\textit{c}$H$\beta$ (i.e., the logarithmic extinction at
H$\beta$) for each nebula. These hydrogen transitions are the strongest and easiest to accurately measure in the nebula spectrum and are fortunately located very close to the other main emission lines used for PN diagnostics.
The observed H$\alpha$/H$\beta$ ratio, when compared to the
recombination value of 2.86 (Aller 1984), gives a logarithmic
extinction at H$\beta$ of:

\begin{equation} \textit{c}(\textrm{H}\beta) = (log(\textrm{H}\alpha /
\textrm{H}\beta) - log(2.86)) / 0.34
\end{equation}


This estimation is based on the relationship between observed and
intrinsic intensities:

\begin{equation}
\frac{I_{obs}(\textrm{H}\alpha)}{I_{obs}(\textrm{H}\beta)} =
\frac{I_{int}(\textrm{H}\alpha)}{I_{int}(\textrm{H}\beta)}
10^{-c(\textrm{H}\beta)[f(\textrm{H}\alpha)-f(\textrm{H}\beta)]},
\end{equation}

where[$\textit{f}$~(H$\alpha$)-$\textit{f}$~(H$\beta$)] = --0.34
from the standard interstellar extinction curve and $\textit{c}$(H$\beta$) is the logarithmic extinction of
H$\beta$. The intrinsic ratio is mildly dependent on temperature:
for T$_{e}$ between 2,500 and 20,000 K, the ratio varies from 3.30
to 2.76 (Osterbrock and Ferland, 2006), although the standard 10,000K is often adopted for T$_{e}$.

The value of $\textit{c}$ and $\textsl{I}$($\lambda$) for all other
lines $\lambda$ was then used to correct for interstellar extinction
using the galactic extinction law from Whitford (1958) in the form
of Miller \& Mathews (1972):
\begin{equation}
I_{cor}(\lambda) = I_{obs}(\lambda) 10^{c[1+f(\lambda)]}.
\end{equation}
Reworking this equation for the H$\beta$ line all other lines could
be corrected for reddening using:
\begin{equation}
\frac{I_{cor}(\lambda)}{I_{cor}(\textrm{H}_{\beta})} =
\frac{I_{obs}(\lambda)}{I_{obs}(\textrm{H}_{\beta})} 10^{cf(\lambda)}.
\end{equation}

Any uncertainty in $\textit{c}$(H$\beta$) will have almost no effect on the
reddening corrected line ratios for H$\beta$/\OIII5007\AA. Even the standard line extinction for \OIII5007\AA~is only
 --0.038 $\textit{f}$~($\lambda$). With such a small correction, the \OIII5007\AA~line flux estimates are very unlikely to be affected by any error estimates in $\textit{c}$(H$\beta$). Errors have been estimated with reference to both
flux calibration and the wide spectral
range between the H$\alpha$ and H$\beta$ lines. This is estimated for H$\alpha$ using the same method
of AAOmega flux calibration described above where the H$\alpha$ spectral line in each
nebula was individually compared to published H$\alpha$ fluxes for the same objects. The maximum estimated uncertainty in
$\textit{c}$(H$\beta$) as a result of combined maximum errors in
line measurement and flux calibration estimations is $<$7\%.


\section{Applied techniques used to identify MCELS PN candidates}
\label{section4}


\subsection{Multi-wavelength and emission-line diagnostic object classification}

The optical spectrum obtained for every potential PN candidate in the MCELS outer LMC sample, including previously known PNe, was carefully measured and analysed in order to assess its classification and probability (see section~\ref{section4.5}). PN candidates were identified by assessing emission-line ratios as described in Reid \& Parker (2006a). Easiest to confirm were those candidates where \NII$>$H$\alpha$ and \HeII4686 was present in emission. These high excitation and likely TypeI (Peimbert 1978) PNe were found at scattered locations across the outer LMC and represent 30$\%$ of the newly discovered PNe. Optical spectra of the remaining candidates were assessed on the basis of \OIII/H$\beta$ ratio where \OIII4959\AA/H$\beta$ $>$3 are more likely to be PNe. For each nebulae, however, an individual assessment must be made where \OIII, \OII, \NII, \SII, \HeII, Ne, Ar, and the Balmer lines are all measured and used for spectral classification. These spectral classifications are conducted hand-in-hand with available optical images using H$\alpha$, SR, \SII~and \OIII~bands in order to exclude possible compact \HII~regions and assess the local environment. This also allowed us to exclude faint objects which were possibly dominated by ambient emission and other sources of contamination. It is worth noting that in the absence of $HST$ imaging, morphological information on LMC PNe is difficult to extract from ground-based data. The best PN candidates were then assessed using the 2MASS $\textsl{J}$, $\textsl{H}$ and $\textsl{K}$ bands and SAGE (Meixner et al. 2006), $Spitzer$~MIR maps as described below.

 The final step was to use the emission line ratio diagnostic plots first introduced by Sabbadin et al. (1977), gradually updated and refined (e.g. Riesgo-Tirado \& L$\acute{o}$pez 2002; Frew \& Parker 2010; Sabin et al. 2011). This important task was undertaken in order to identify and exclude interlopers which have biased and effected most previous PN catalogues, both in our own Galaxy and in the Magellanic Clouds. Contamination by the many types of possible PN mimics may adversely affect the true form of the PNLF. This is particularly important at the bright end whose truncated exponential fit is used as an extragalactic distance indicator (Ciardullo et al, 2002; Reid \& Parker 2010).

\subsection{Multi-wavelength identity checks}

A combined multi-wavelength and spectral analysis is a very complete and robust method of assessing the veracity of PNe. It is clear, however, that an optical spectrum and high resolution, deep optical image are the best starting point for the identification of a PN. It is additionally helpful to examine the NIR bands using 2MASS (Skrutskie et al. 2006, Cutri et al., 2003) or the new Vista Magellanic Cloud (VMC survey; Cioni et al, 2011) $\textsl{Y}$, $\textsl{J}$ and $\textsl{K}$$_{s}$ colours in order to identify the increased reddening, expected in most PNe. SAGE (Meixner et al. 2006) observations in the MIR provide much needed information regarding the presence of Polycyclic Aromatic Hydrocarbons (PAHs), dust and other thermal emissions which help to separate PNe from emission-line stars and \HII~regions. Indeed, Parker et al (2012) have recently shown it is possible to identify high-quality PN candidates purely on the basis of their MIR characteristics.

The brighter PNe candidates in the new LMC sample were crosschecked against 2MASS NIR images and photometric data at $\textsl{J}$(1.25\,$\mu$m), $\textsl{H}$(1.65\,$\mu$m) and $\textsl{K}$$_{s}$(2.16\,$\mu$m) band-passes in order to identify and separate bright stellar sources and large emission objects from PNe. If a stellar continuum is dominating over and above what is expected from a PN central star with or without NIR emission lines, the flux will be relatively brighter towards the shorter NIR wavelengths. If the object is an extended emission source, the warm dust will be much stronger at longer wavelengths. Since a PN spectrum at each of these bands may be composed of various mixtures of emission lines and/or continuum, only objects with a large excess in either $\textsl{J}$ or $\textsl{K}$ are likely to be non-PNe. Combining these data with MIR data may further assist in the classification process.

\begin{figure}
\begin{center}
  \includegraphics[width=0.48\textwidth]{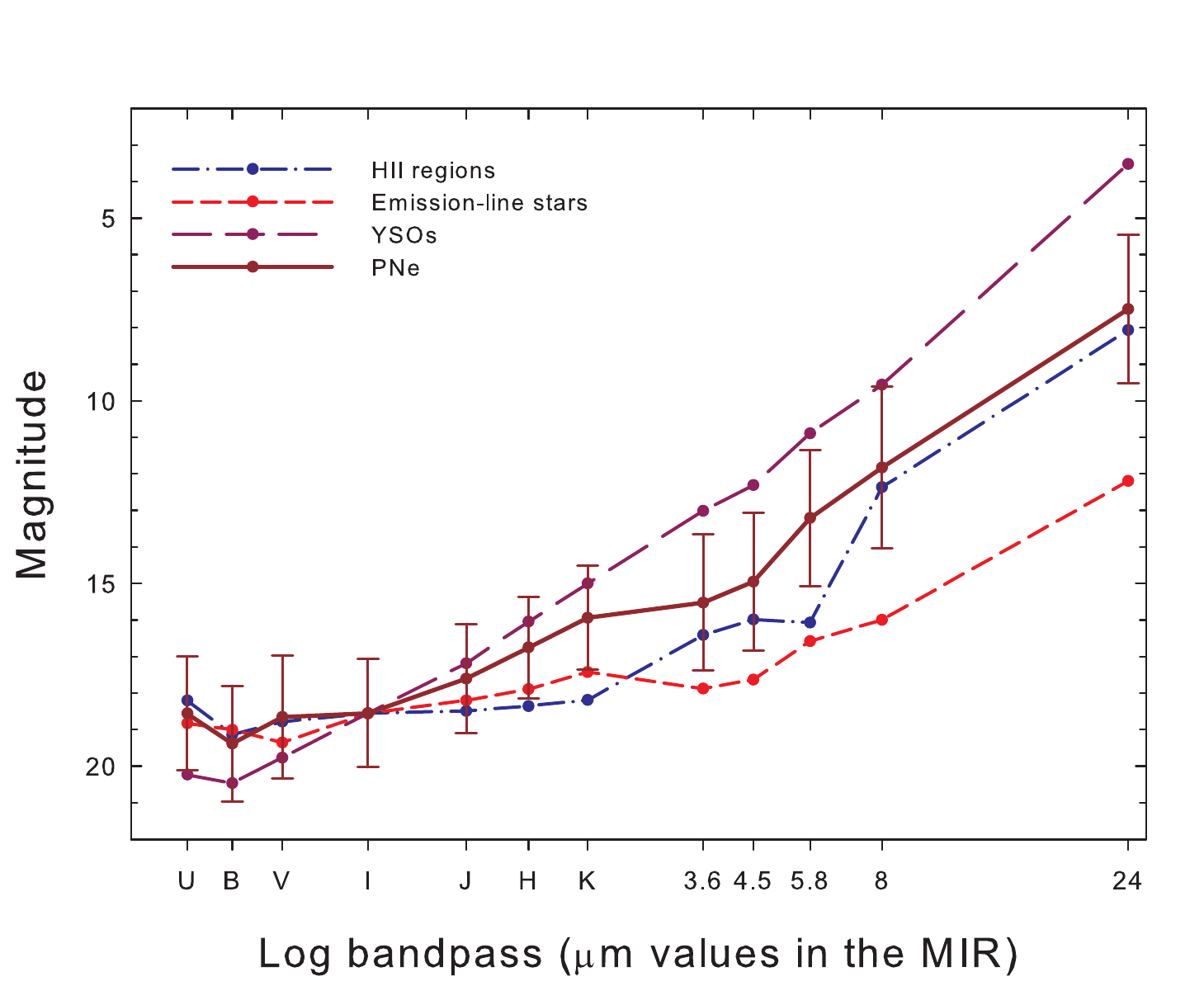}\\
  \caption{The mean SEDs covering $U$~band to 24\,$\mu$m in the MIR for \HII~regions (dark blue dash-dot), Emission-line stars (red short-dashed) and YSOs (purple long-dashed). The average fluxes for each object type are derived solely from LMC objects. These are the object types most commonly confused with PNe. They have SEDs that correspond closely with PNe. The SED based on confirmed PNe in the LMC is indicated by the solid dark brown line where the error bars give the standard deviation. Objects are normalized to the mean PN magnitude of 18.85 in the $\textsl{I}$~band.}
  \label{Figure5}
  \end{center}
\begin{center}
  \includegraphics[width=0.48\textwidth]{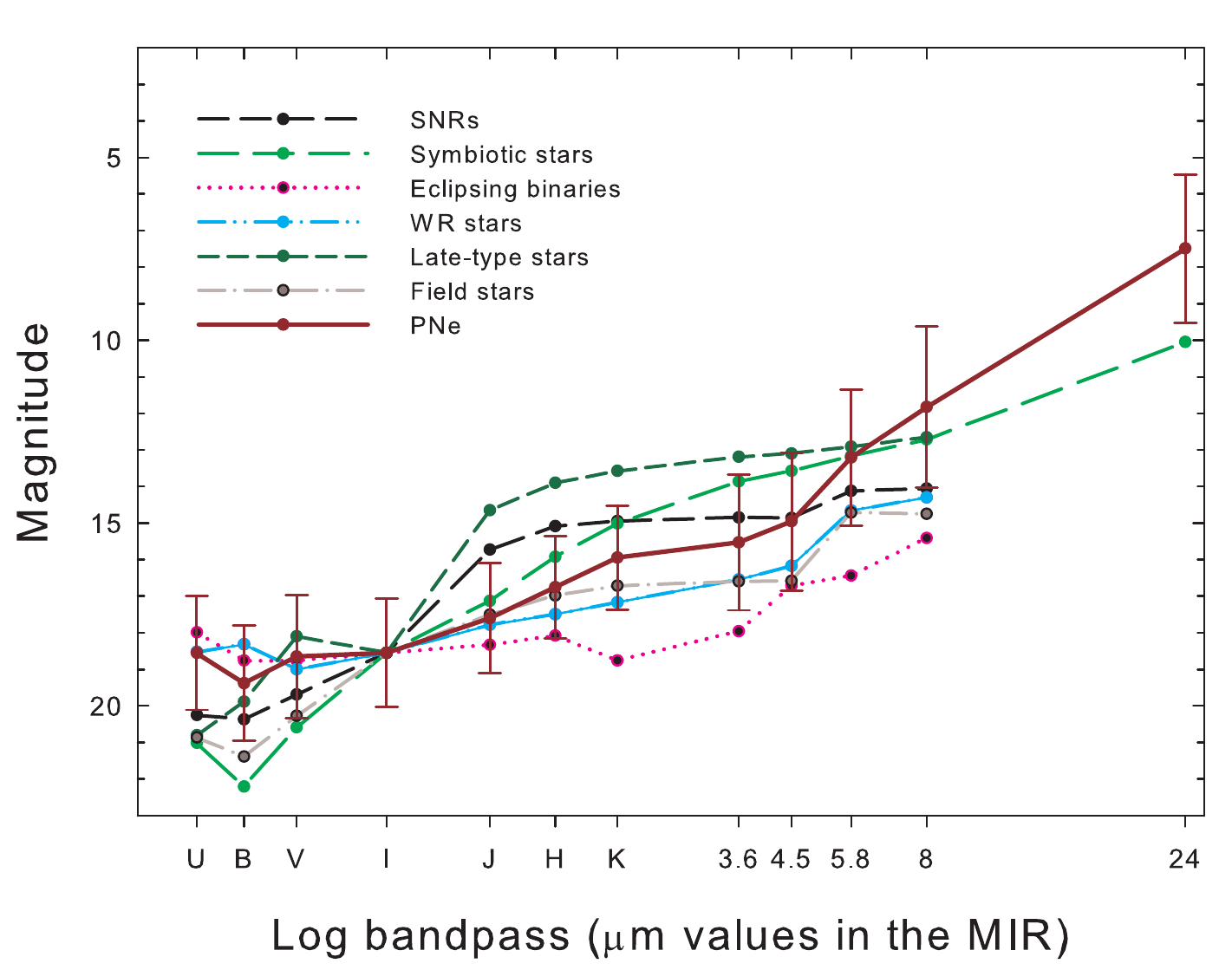}\\
  \caption{Same as above but for SNRs (black dashed), symbiotic stars (light green long-dashed), Eclipsing binaries (pink dotted), WR stars (light blue dash-dot), late-type stars (dark green short-dashed), field stars (grey dash-dot) and all PNe in the LMC (dark brown solid). The SEDs formed by these objects allow them to be clearly separated from PNe.}
  \label{Figure6}
  \end{center}
  \end{figure}

\begin{figure}
\begin{center}
  \includegraphics[width=0.48\textwidth]{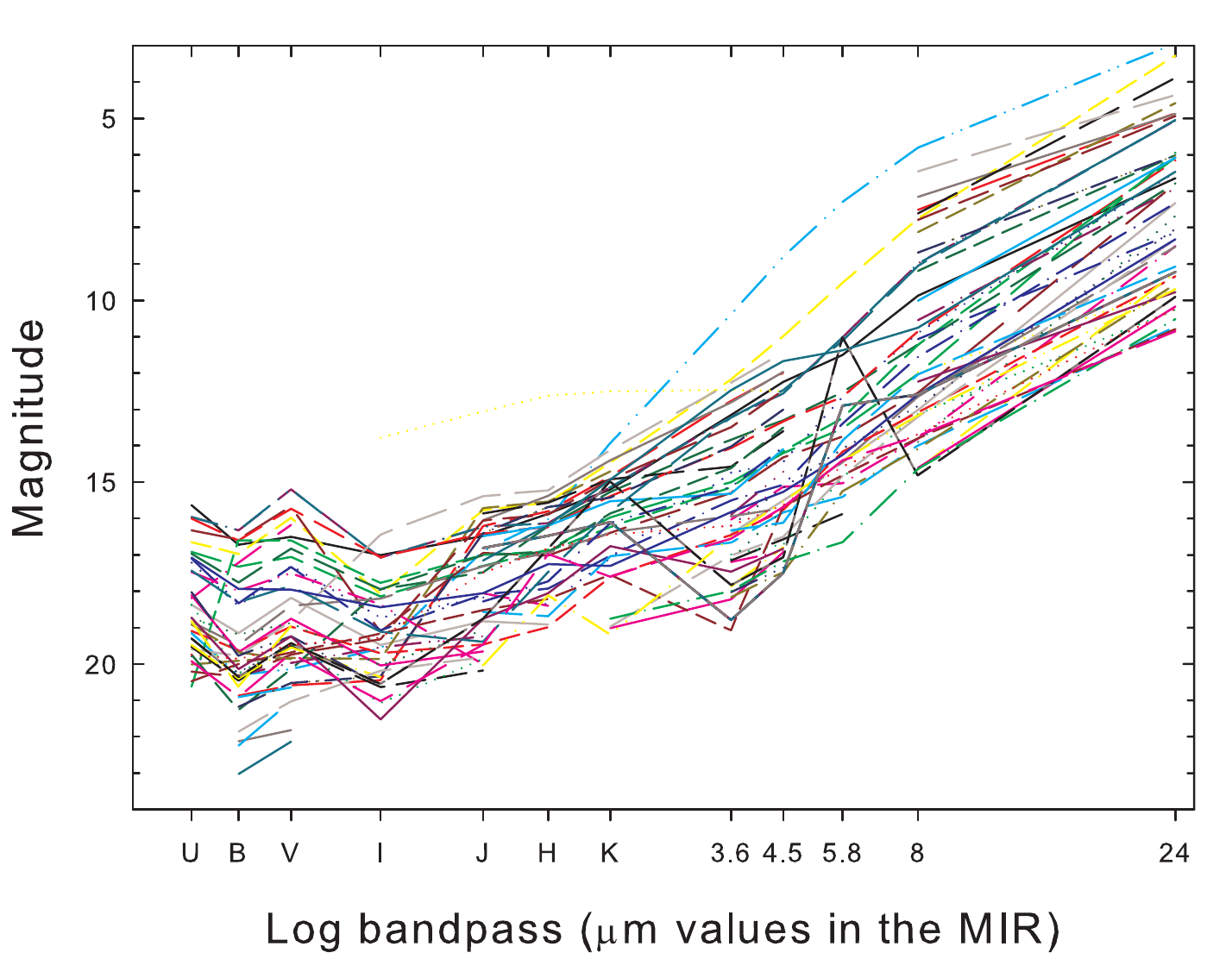}\\
  \caption{The SEDs covering $U$~band to 24\,$\mu$m in the MIR for all previously known PNe in the outer LMC where data linking at least two consecutive points are available.}
  \label{Figure7}
  \end{center}
\begin{center}
  \includegraphics[width=0.48\textwidth]{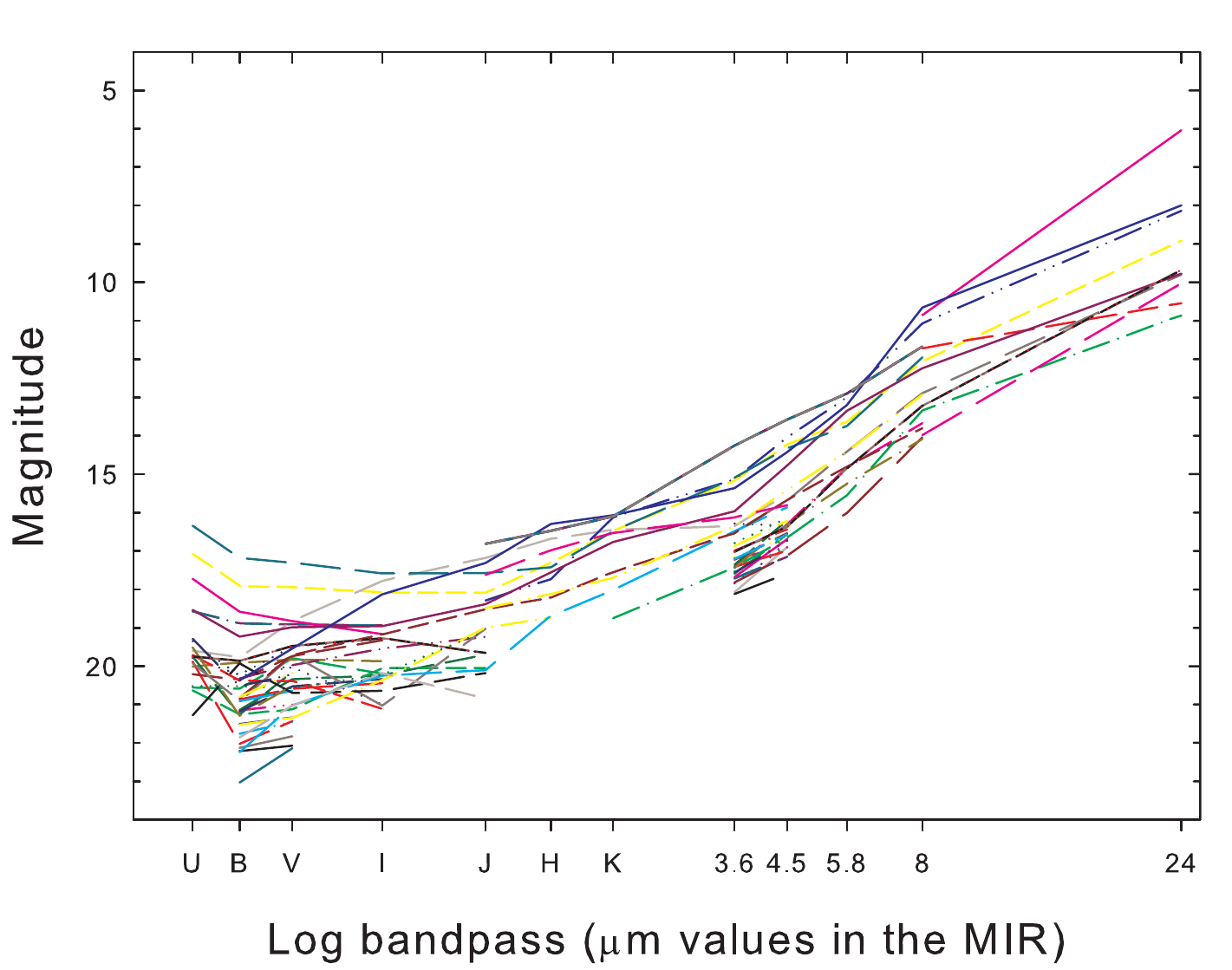}\\
  \caption{The SED covering $U$~band to 24\,$\mu$m in the MIR for all newly identified PNe in the outer LMC where data linking at least two consecutive points is available.}
  \label{Figure8}
  \end{center}
  \end{figure}

By overlaying SAGE false colours at 3.6, 4.5, 5.8 and 8\,$\mu$m as well as supplementary $WISE$ (Wright et al. 2010) bands at 3.4, 4.6$\mu$m, we could visually assess the expected IR excess with decreasing wavelength. $WISE$ has lower resolution than SAGE but additionally provides wavelength bands at 12 and 22$\mu$m, providing extra confirmation at longer wavelengths. Magnitudes were taken from the published $Spitzer$ Infrared Array Camera (IRAC; Fazio et al. 2004) SAGE data from Hora et al. (2008) and from Gruendl and Chu (2009). The application of false colours allowed us to observe and compare both luminosity ratios and angular extent at each waveband. Using the extracted NIR-MIR magnitudes, diagnostic plots, for all possible PNe were created with the aid of this multi-wavelength data (e.g. Hora et al. 2008). This allowed us to reduce the number of candidate PNe from an initial 87 to the 61 presented here. Included in these plots were 3,000 LMC stars which largely do not support extended atmospheres or strong emission lines. By comparing 11 different combinations of NIR and MIR wavelengths against each other, in general the PNe cleanly separated from the stellar population. PN candidates that consistently fell within the region occupied by stars in key plots were re-assessed both spectroscopically and for their local environment. Unless it was clear that candidates were well evolved, and the dusty envelope had dissolved and dispersed (see examples in Hora et al., 2008), the candidate was dropped from initial inclusion in the present PN list. The individual results of this multi-wavelength analysis, covering the whole LMC (as well as these newly identified candidates in the outer LMC) is presented in the sister paper to this one (Reid \& Parker, 2013).

In addition to an analysis of the false colour images, constructed spectral energy distributions (SEDs) covering $\textsl{U}$ to 24\,$\mu$m allowed us to compare our newly identified PN candidates with previously known PNe in the outer LMC. Magnitudes for $\textsl{U}$,$\textsl{B}$,$\textsl{V}$ and $\textsl{I}$ were taken from the Magellanic Clouds Photometric Survey (Zaritsky et al. 2004), NIR magnitudes are from 2MASS (Cutri et al., 2003) and IRSF (Kato et al., 2007). The SEDs were constructed using raw data, uncorrected for extinction. We first constructed a comparison plot of SEDs for PN mimics in the LMC in order to compare the shape of their SEDs with that of PNe. Such plots can assist in identifying non-PNe since emission nebulae containing dust components, emission lines and a central stellar source create a characteristic shape. Other objects compared include compact \HII~regions, compact bubbles which may be part of a SNR, symbiotic stars, eclipsing binaries, Wolf-Rayet (WR) stars, emission-line stars, young stellar objects (YSOs), late-type stars and ordinary field stars. Each object type from the LMC was averaged with the resulting mean for each bandpass shown in Figs.~\ref{Figure5} and~\ref{Figure6}. The average SED for LMC PNe includes all 715 known and new PNe, including those presented here for the first time, and represents the true mean flux for LMC PNe at each band. Error bars show the standard deviations found for PNe, since they are of principal interest.

Similarity in the overall shape of SEDs for various emission sources has led to PN mimics being included in PN catalogues over past decades (see Frew \& Parker 2010). Fig.~\ref{Figure5} compares the averaged SEDs for PNe against those derived for \HII~regions, emission-line stars and YSOs, the objects most commonly mistaken for PNe. All mean object types have been normalized to the mean I-band magnitude of 18.85 for LMC PNe. Although similar, there are some interesting differences in these SEDs that can assist in identification. YSOs show the greatest luminosity range from the optical to the MIR. They increase with particular strength through the NIR region, almost forming a straight line from $\textsl{B}$ to 24$\mu$m. In contrast, \HII~regions show a somewhat flat SED extending from $\textsl{V}$ across to $\textsl{K}$ in the NIR. After a rise to the MIR they are also rather flat until 8$\mu$m, where they brighten by an average 4 mag. Emission-line stars follow the SEDs of \HII~regions closely until 8$\mu$m, after which a clear lack of PAHs means they only continue a gradual rise driven by dust continuum. Only five emission-line stars in our sample of the 970 with IRAC data have a published flux at the 24$\mu$m band. This is the result of both the sensitivity cut-off for Spitzer MIPS in the LMC and a general lack of available MIPS cross-matched data. The SED transition from 8$\mu$m to 24$\mu$m should therefore be accepted with some caution as it may only represent the brightest emission-line stars in the LMC. The subtle differences between the SEDs of these objects and PNe can be of enormous benefit when low resolution optical spectra are inconclusive.

In Fig.~\ref{Figure6} we show the mean SEDs of several object types that are less often confused with PNe due to strong emission-line characteristics in the optical. In terms of their SEDs, the closest counterparts to PNe are the WR stars and symbiotic stars. WR stars are brighter in $\textsl{B}$ due to a strong ionized helium component. Symbiotic stars increase in luminosity red-ward from $\textsl{B}$, presumably due to the interaction of a late-type stellar companion. Late-type stars have the largest deviation, also extending to brighter magnitudes in the NIR than any other object types. Although these cool stars resemble PNe in H$\alpha$/red subtracted images, an optical spectrum can instantly identify these stars due to the rising red continuum and TiO bands.

Since all the plotted objects are located in the LMC at the same average distance, we can directly compare mean luminosity levels. PNe have an optical luminosity that is on average comparable to field stars, YSOs and SNRs. Late-type stars and symbiotic stars have comparable luminosities to PNe only in the $\textsl{U}$ and $\textsl{B}$ bands. Their luminosities increase substantially through the $\textsl{B}$ and $\textsl{V}$ bands to become the brightest sources to be found in the NIR. Unlike PNe, however, they do not experience a substantial rise in luminosity through the MIR, indicating a low dust and molecular content. Although the SEDs of PNe also rise towards the NIR, the rise is more modest, placing them among the faintest of objects across the NIR bands. PNe gain in luminosity across the MIR, somewhat matching the rise experienced by WR stars and \HII~regions, although the later may experience more irregular variations depending on the mass, density and ionizing source of the gas. Those objects, brightest at optical wavelengths such as emission-line stars, WR stars, eclipsing binaries and \HII~regions may experience a slight decrease in luminosity towards the $\textsl{V}$ band but in most cases will experience a modest rise across the NIR and into the MIR. Within the MIR they each experience a rise, notably with WR stars increasing substantially between 4.5 and 5.8$\mu$m and \HII~regions doing the same between 5.8 and 8$\mu$m. Although all of these SEDs are quite similar there are key differences at certain wavelengths and so they can be of some use in weeding out, or at least identifying, suspicious PN candidates.

The plots, shown in Fig.~\ref{Figure7} for the previously known PNe in the outer LMC and Fig.~\ref{Figure8} for the newly identified PNe in the same outer regions provide an additional certainty as to the new classification of these objects. At this stage the plots are shown in order to verify and compare the general trends for LMC PNe rather than assess individual objects. For that reason we have not identified PNe that show somewhat unusual behaviour.

In both plots the PNe SEDs can be defined as showing an almost flat or downward slope from the blue towards the $\textsl{I}$~band followed by a steady rise towards 24$\mu$m. A noticeable rise in the SEDs for previously known PNe around the $\textsl{V}$~band shows the strength of the \OIII~emission lines in these brighter objects compared to the more highly evolved and fainter PNe found in our new sample which are generally stronger in their \NII~emission. The steady rise in the SEDs of both previously known and newly discovered PNe is evidence that the cool, warm, and hot dust has an equal effect on the SEDs both while PNe are young and when they are optically faint and evolved. Presumably atomic and molecular emission lines will still be present at some level in the faint PNe. A number of PNe show some variation in the SEDs moving from the $\textsl{I}$-band up to 8\,$\mu$m but this may be due to either variations in the relative strengths of the stellar continuum affecting the warm dust emission, scattered light around the NIR, H$_{2}$ emission, unusual variations in the PAH emission at 5.8 and 8\,$\mu$m and/or forbidden line emission from ionised gas at discrete wavelengths that could affect band strengths. Such an example is the MIR \OIV~emission line which is a proxy for the \HeII~line in the optical and so arises within high excitation PNe. Where data are available, we find that all the PNe show a rise between 8 and 24$\mu$m, indicating a central star that is hot enough to produce a warm dust component.

One issue for 2MASS data in the LMC is its sensitivity cutoff. Although the 2MASS sensitivity at S/N = 10 is already reached at 15.8, 15.1 and 14.3 mag for $\textsl{J}$, $\textsl{H}$ and $\textsl{K$_{s}$}$ respectively, some strong emission lines which often dominate these bands in PNe can still be detected in the LMC sample. The brightest in the $\textsl{J}$ band is Paschen $\beta$ although a spectrum will also reveal the presence of \HeI, \FeII~and \OI. The $\textsl{H}$ band can be dominated by the Brackett series with \HeI\,1.7002\,$\mu$m and occasionally \FeII\,1.6440\,$\mu$m. The brightest line in the $\textsl{K$_{s}$}$ band is Brackett $\gamma$ with \HeI\,2.058, 2.112\,$\mu$m. Although it is useful to know that these lines may be present, we know that they may be almost missing in young PNe where a strong warm dust continuum may dominate (Hora et al, 1999). Fig.~\ref{Figure8} shows us that these lines are also likely to be either missing or below the detection limit in highly evolved PNe which include most of the newly discovered PNe in our sample.

\subsection{Diagnostic plots}

With reliable flux calibration of the LMC PN observational data covering almost every PN in the LMC, basic diagnostic plots provide an overall snapshot of line characteristics and ratios. These plots can be useful for identifying PN mimics (Frew \& Parker 2010). However, considerable overlap between the regions occupied by PNe, \HII~regions, symbiotic stars, WR stars, SNRs, AGNs and emission-line stars of various types means that the identification of suspicious, outlying objects will have to be referred back to earlier forms of identification. In other words, they will rely more heavily on multiple wavelength analysis and high resolution optical imaging.

\begin{figure}
\begin{center}
  \includegraphics[width=0.485\textwidth]{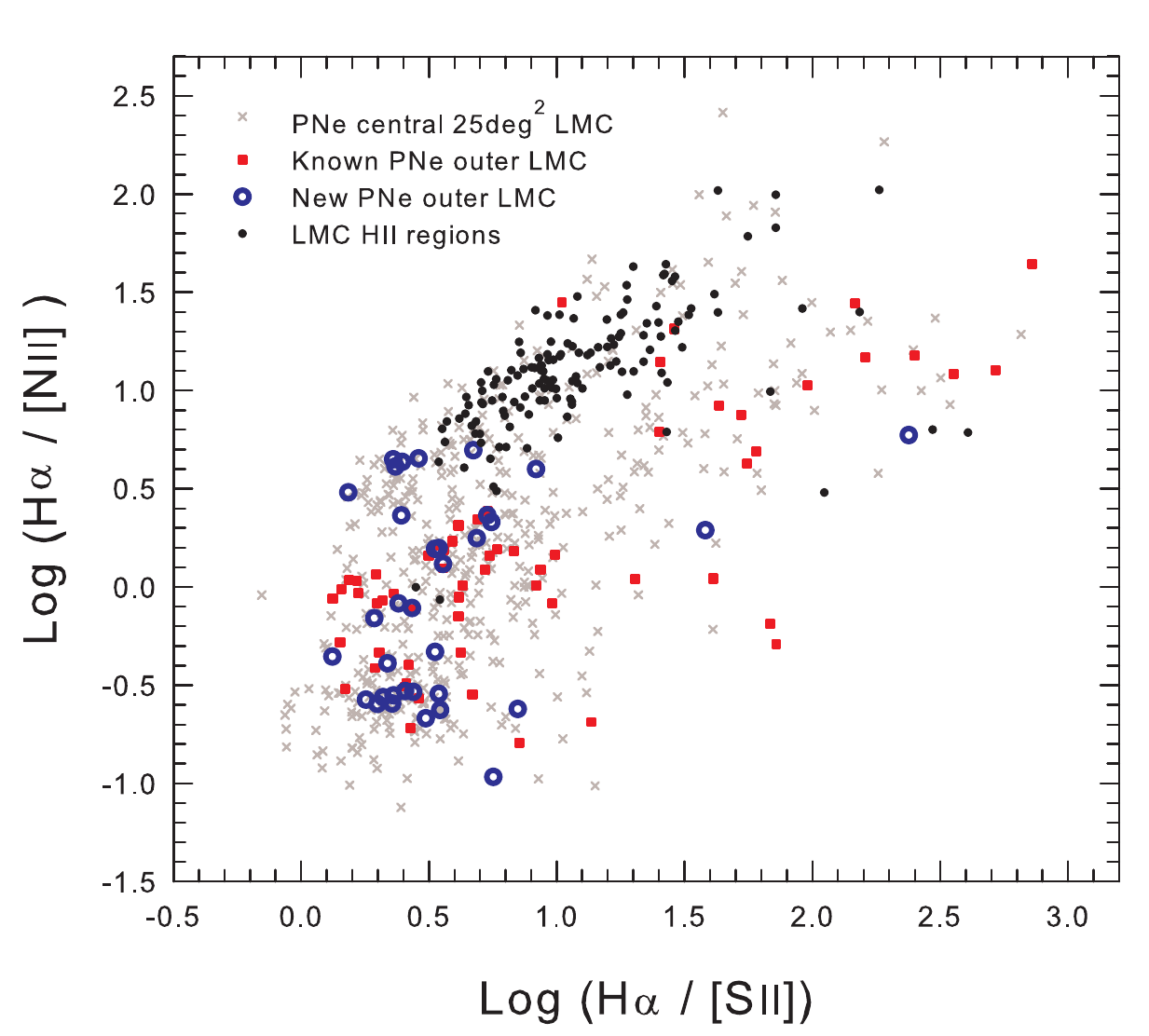}\\
  \caption{The log (H$\alpha$/\NII6584+6548\AA) versus log (H$\alpha$/\SII6716+6731\AA) diagnostic diagram. PNe within the central 25deg2 main bar region of the LMC are marked as grey crosses, previously known PNe in the outer LMC are marked as red squares and the newly identified PNe in the outer LMC are marked as open blue circles. \HII~regions are shown as filled black circles.}
  \label{Figure9}
  \end{center}
  \end{figure}
  \begin{figure}
\begin{center}
  \includegraphics[width=0.48\textwidth]{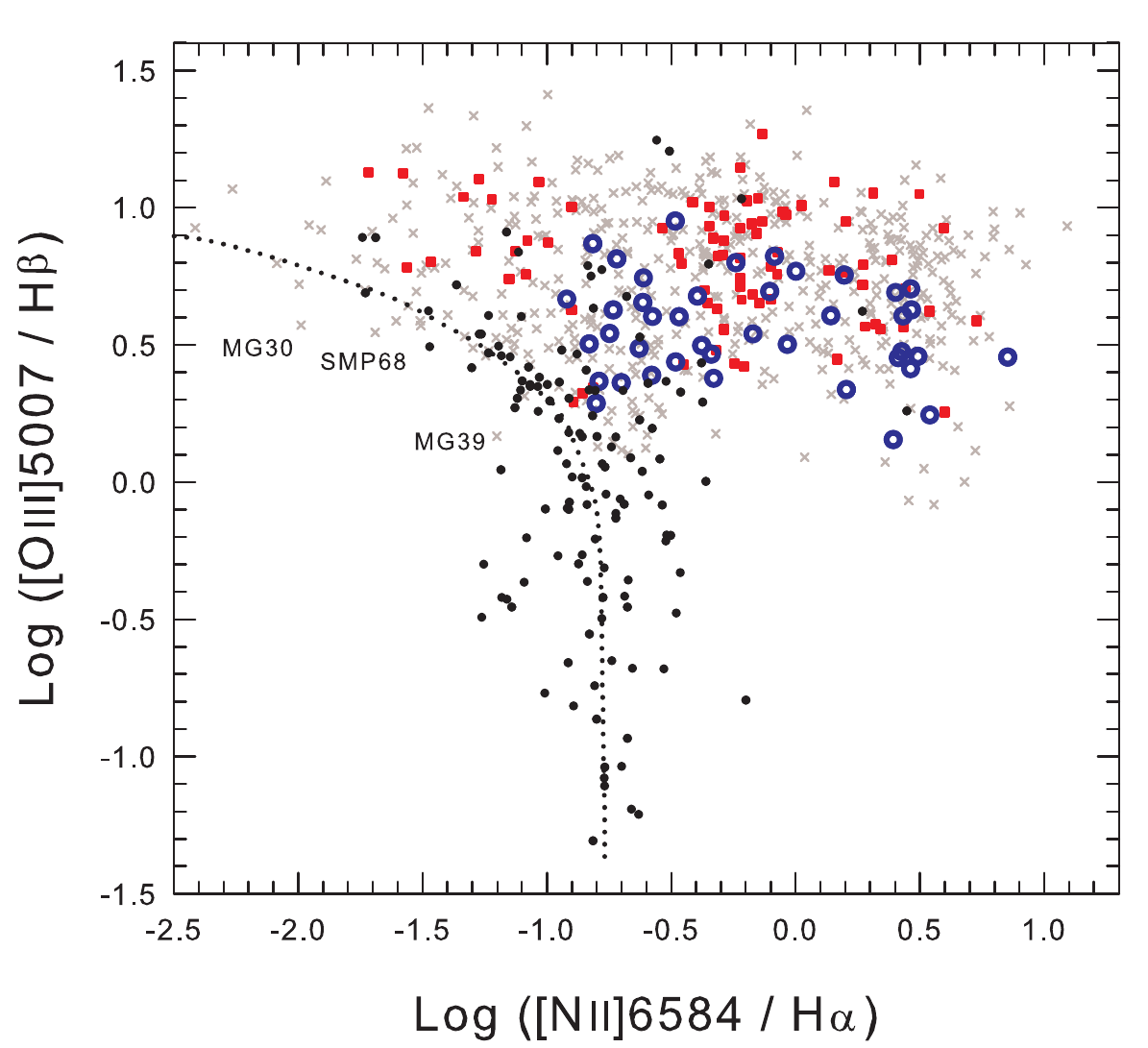}\\
  \caption{The log (\OIII5007\AA/H$\beta$) versus log (\NII6584\AA/H$\alpha$) diagnostic diagram. The colours and symbols have the same representation as defined in Fig.~\ref{Figure9}. The black broken line represents the line of best fit for the included \HII~regions in the LMC.}
  \label{Figure10}
  \end{center}
\begin{center}
  \includegraphics[width=0.48\textwidth]{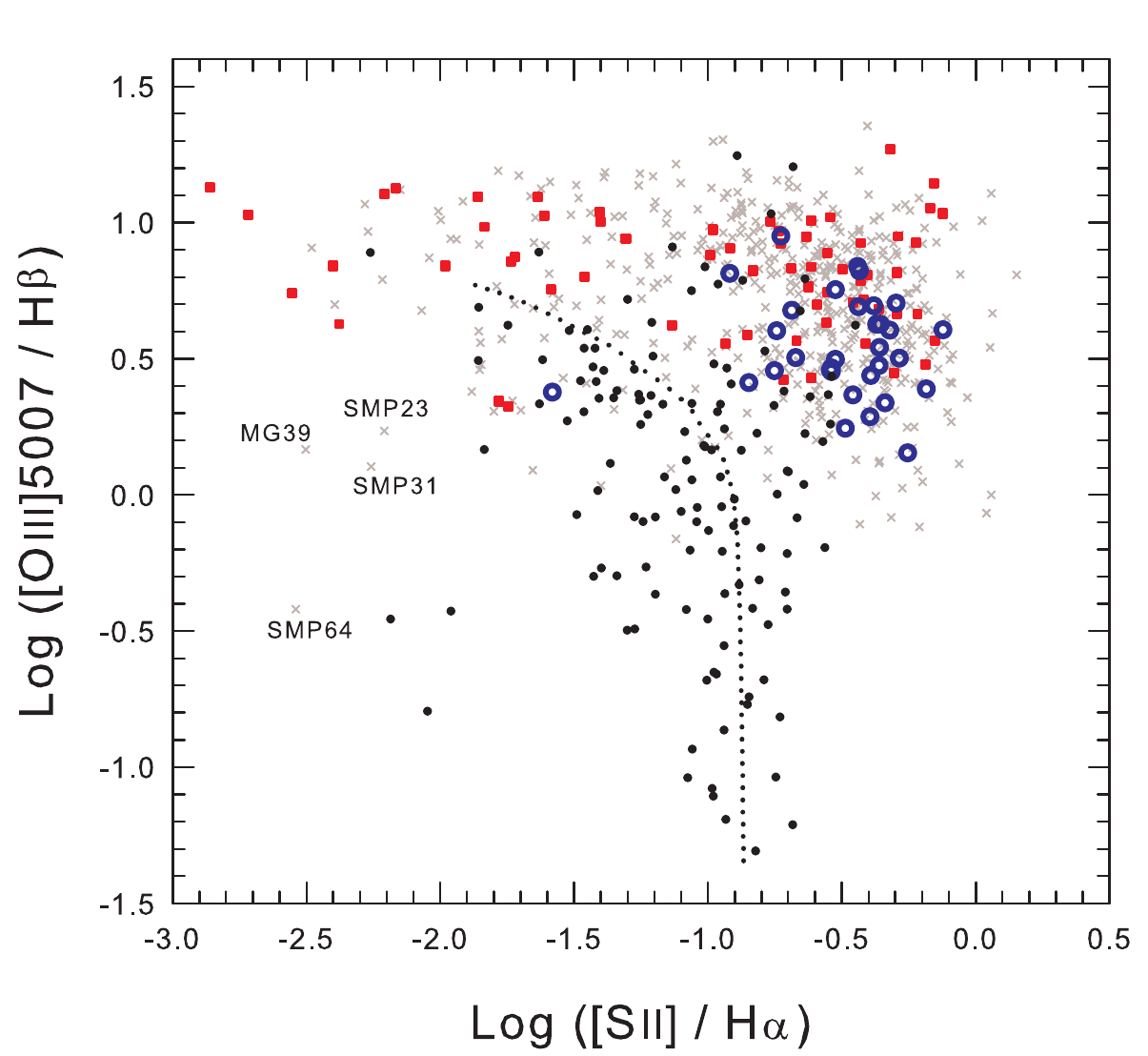}\\
  \caption{The log (\OIII5007\AA/H$\beta$) versus log (\SII6717+6731\AA/H$\alpha$) diagnostic diagram. Symbols and colours are the same as for Fig.~\ref{Figure8}.}
  \label{Figure11}
  \end{center}
  \end{figure}

Our low and medium resolution spectra for all the LMC PNe provide the fluxes to plot log H$\alpha$/\NII~versus log H$\alpha$/\SII. Since the subtracted line ratios on each axis are at similar wavelengths, the reliance on accurate flux calibration and reddening errors is substantially minimized. Fig.~\ref{Figure9} includes all PNe in the LMC where lines were able to be measured at greater than 2$\sigma$ above the noise level. Only 34 out of the 61 (55$\%$) newly identified PNe in the outer LMC have measurable \NII6584+6548\AA~and \SII6716+6731\AA~lines allowing them to be shown as open blue circles in Fig.~\ref{Figure9}. This number compares to 62 of the 98 (63$\%$), brighter, previously known PNe in the outer LMC (red squares in Fig.~\ref{Figure9}) where both the \NII6584+6548\AA~and \SII6716+6731\AA~emission lines are able to be measured. Such a lack of identifiable \NII~and/or \SII~emission is not uncommon and is often the case for very high or very low excitation PNe. The classification of such PNe requires a heavier reliance on imaging, other emission lines and multi-wavelength data. Since the \SII~lines are often enhanced by shock excitation (Dopita et al. 1995) the lack of these lines in the faint new outer LMC PNe is an indication of very low excitation and lack of substantial ambient emission in the ISM.

Fig.~\ref{Figure9} shows a large variation in the relative \NII~line strength from log 2.4 to log -1.1. Those PNe grouped at the lower left of the plot, around -0.5($\textsl{y-axis}$) and 0.4($\textsl{x-axis}$) -0.5 : 0.4 hereafter, are likely to be Type I PNe (Kingsburgh \& Barlow, 1994) as they are highly enriched in nitrogen. If helium is present at the abundance level He/H $>$ 0.10 and N$^{+}$/O$^{+}$ $>$ 0.25, the PN can definitely be classed as TypeI according the average metallicity of the LMC (Leisy \& Dennefeld 2006). Opposed to these types are the low excitation PNe which are seen from $\sim$0.5 : 0.5 to 2.0 : 2.0, following a track where ionization ratios are more or less equal. It is along this track and above that we expect to find most of the LMC \HII~regions. In Fig.~\ref{Figure9}, the known LMC \HII~regions, measured and flux calibrated in the same manner as that used for PNe, are shown as black filled circles. Although the majority of them occupy the region of the plot containing low excitation PNe, a number of them can also be found in areas occupied by medium-to-high excitation PNe. Some of these were re-classified from PNe to \HII~regions using SEDs based on VMC and other multi-wavelength data where available. These gas formations are very useful since they trace the metallicity of the host galaxy with more precision than PNe.

The log (\OIII5007\AA/H$\beta$) versus log (\NII6584\AA/H$\alpha$) plot is shown in Fig.~\ref{Figure10} together with the log (\OIII5007\AA/H$\beta$) versus log (\SII6717+6731\AA/H$\alpha$) diagnostic diagram in Fig.~\ref{Figure11}. These plots are somewhat more useful in separating PNe from \HII~regions, at least where the \OIII/H$\beta$ ratio is high for PNe and low for \HII~regions. PNe group in the upper right of the plot about a loci of log (\OIII5007\AA/H$\beta$) versus log (\NII6584\AA/H$\alpha$) 0.7 : -0.5. In the mid-range, from log \OIII5007\AA/H$\beta$ $\sim$0-0.7 there is considerable overlap. Many well-confirmed PNe such as MG30, MG39, SMP23, SMP31, SMP64, SMP68 lie in the area beyond the main division marked by \HII~regions.

Unfortunately, in the LMC there are a number of very large, high excitation \HII~regions and giant superbubbles, powered by OB associations which can cause elevated \SII/H$\alpha$ ratios through the action of stellar winds (eg. Lasker 1977). This may account for some of the high \SII/H$\alpha$ ratios seen here in Fig.~\ref{Figure11}. The other factor to consider is that most of the \HII~regions included in these plots are very compact PN mimics, many of which have an angular extent not much larger than that of confirmed PNe. Only a heavy reliance on multi-wavelength analysis has enabled them to be separated them from true PNe. The purified list of newly identified MCELS PNe are presented below.

\subsection{The newly confirmed PN candidates}
\label{section4.5}

The newly identified and spectroscopically confirmed MCELS PN candidates are shown in Table~\ref{table 1}. Column 1 gives the newly assigned RP reference number, Columns 2 and 3 give the J2000 positions in RA and DEC respectively accurate to $\sim$2 arcsec (as checked against the online SuperCOSMOS R positions), Column 4 gives our PNe probability ranking of `True', `Likely' or `Possible' based on the process adopted for our earlier PN catalogue. Any special comments regarding individual objects are given in Column 5.

Included in this outer LMC area covered by MCELS are 106 previously known PNe, all of which were independently re-identified and included in our spectroscopic followup. All showed standard PN emission lines with the exception of Mo22. In the published position for this PN at RA: 05 21 26.6 Dec: -62 34 12.5, (see Leisy \& Dennefeld 2006) we found an ordinary field star with a magnitude of B=17.9. There is no H$\alpha$ excess visible and no emission found in the stellar spectrum. We have searched the local area in case of an astrometric error, however, there are no discrete emission sources nearby. Consequently, we re-classify this object as `not a PN'. All of the 61 new spectroscopically confirmed PNe were first discovered as candidates from the MCELS maps. They were then cross-checked against our UKST LMC SuperCOSMOS maps and other extant wide field images in the 2MASS and VMC NIR surveys and from the equivalent SAGE and $WISE$~surveys in the MIR. Following on from this work, a detailed multi-wavelength study (including all the previously known and newly discovered PN in the LMC) was undertaken and is presented in the sister paper released simultaneously with this work (Reid \& Parker, 2013).

In the Appendix we provide thumbnail cut-out images for each PN candidate as they were found in the MCELS merged three-colour maps and then the equivalent single exposure UKST H$\alpha$ and SR merged colour images. PNe RP4489 and 4687 were uncovered in the UKST images, just outside of the MCELS area so no MCELS image is provided for them. They have been included here to demonstrate the potential of these deep UKST images. To the right of these discovery images we provide confirmatory spectra where the blue and red arm data have been spliced together and flux calibrated to cover the full optical wavelength range. Most spectra are from our AAOmega blue and red arm observations. Where spectra were obtained using 6dF on the UKST or from long slit observations on the SAAO 1.9 m we have made a special note on the image itself.

\begin{table}
\begin{center}
\caption{Newly identified PN candidates in the LMC. Shown are the RP reference number in Column 1, the J2000 RA and Dec in Columns 2 and 3, the probability:- `T' = true, `L' = likely, `P' = possible- in column 4 and comments in column 5. In the `Comments' Column, c= circular, f= faint, e= elliptical, ext= extensions, near star = the PN candidate is very close to or difficult to separate from a stellar source. }
\begin{tabular}{lllcl}
\hline\hline \noalign{\smallskip}
RP cat. & RA $^{(h~m~s)}$ &  Dec $^{(\circ~_{'}~_{''})}$   & Prob. & Comments  \\
     No.   &  J2000   &  J2000        &          &       \\
  \hline\noalign{\smallskip}
RP1870	&	05 01 59.65	&	-68 02 35.47	&	T	& c	\\
RP1877	&	04 56 01.03	&	-67 54 42.84	&	L	& e	\\
RP2347	&	04 35 43.63	&	-70 33 15.50	&	T	& f	\\
RP2419	&	04 40 33.37	&	-69 05 33.56	&	T	& c,f	\\
RP2547	&	04 46 11.10	&	-70 01 32.40	&	T	& c,f	\\
RP2637	&	04 49 27.25	&	-68 13 03.80	&	T	& c	\\
RP2698	&	04 51 30.59	&	-69 36 50.50	&	T	& e	\\
RP2708	&	04 51 47.59	&	-71 33 03.70	&	T	& c	\\
RP2784	&	04 53 43.69	&	-66 37 42.78	&	T	& c	\\
RP2862	&	04 56 00.88	&	-67 54 44.40	&	T	& c	\\
RP2914	&	04 56 56.23	&	-65 07 44.50	&	T	& c, near star	\\
RP3228	&	05 07 46.92	&	-65 59 43.40	&	L	& e,f	\\
RP3308	&	05 10 23.90	&	-64 20 26.84	&	T	& c,f	\\
RP3332	&	05 13 52.21	&	-66 24 57.02	&	T	& c, near star	\\
RP3334	&	05 14 29.62	&	-71 37 44.50	&	T	& c	\\
RP3335	&	05 14 34.95	&	-65 00 18.81	&	T	& c	\\
RP3343	&	05 16 37.18	&	-71 52 50.70	&	L	& c	\\
RP3353	&	05 17 37.39	&	-71 59 53.20	&	T	& c, near star	\\
RP3367	&	05 20 36.16	&	-65 26 26.40	&	L	& c, ext?	\\
RP3368	&	05 20 42.37	&	-65 26 05.10	&	L	& c, near star	\\
RP3401	&	05 23 19.75	&	-71 48 21.18	&	T	& c	\\
RP3409	&	05 23 28.79	&	-65 20 13.10	&	T	& e	\\
RP3425	&	05 24 09.70	&	-65 33 08.21	&	T	& e	\\
RP3426	&	05 23 00.63	&	-65 37 47.08	&	P	& c	\\
RP3432	&	05 24 19.87	&	-64 31 13.37	&	P	& e, near star	\\
RP3449	&	05 24 41.17	&	-72 01 45.00	&	T	& c, bright	\\
RP3464	&	05 24 57.79	&	-66 05 59.70	&	T	& e	\\
RP3520	&	05 26 11.84	&	-71 28 55.40	&	P	& c	\\
RP3533	&	05 26 23.96	&	-71 27 50.60	&	P	& e	\\
RP3543	&	05 26 36.16	&	-71 27 00.60	&	P	& c, near star	\\
RP3576	&	05 26 55.82	&	-71 27 55.90	&	P	& c, f	\\
RP3601	&	05 28 55.69	&	-72 17 40.76	&	T	& c	\\
RP3661	&	05 28 01.07	&	-65 34 39.70	&	T	& c, ext?	\\
RP3808	&	05 33 03.00	&	-66 18 29.00	&	T	& c	\\
RP3836	&	05 34 21.20	&	-66 25 25.00	&	T	& c	\\
RP4029	&	05 44 30.44	&	-71 36 03.80	&	T	& e, f	\\
RP4034	&	05 45 29.23	&	-72 18 43.50	&	T	& c	\\
RP4041	&	05 46 00.10	&	-72 29 31.00	&	T	& c	\\
RP4049	&	05 46 24.36	&	-66 02 49.50	&	T	& e	\\
RP4062	&	05 47 22.89	&	-71 33 13.50	&	L	& c,f, near star	\\
RP4065	&	05 47 27.53	&	-65 01 50.80	&	T	& c	\\
RP4078	&	05 48 40.74	&	-66 48 04.20	&	T	& c	\\
RP4080	&	05 58 15.81	&	-71 12 55.82	&	T	& e?	\\
RP4081	&	05 48 42.52	&	-71 09 39.80	&	T	& c,f	\\
RP4139	&	05 52 34.86	&	-71 52 06.60	&	T	& c	\\
RP4143	&	05 52 49.34	&	-69 07 08.20	&	T	& e	\\
RP4146	&	05 53 02.81	&	-70 54 34.60	&	T	& e, near star	\\
RP4151	&	05 53 26.34	&	-70 59 13.80	&	T	& c	\\
RP4176	&	05 54 35.23	&	-64 58 15.80	&	T	& e,f, near star	\\
RP4193	&	05 55 29.61	&	-65 03 38.90	&	L	& c,f	\\
RP4197	&	05 55 48.48	&	-64 57 00.80	&	T	& c,f	\\
RP4207	&	05 56 12.79	&	-68 49 00.30	&	T	& c	\\
RP4210	&	05 56 30.56	&	-65 35 48.80	&	T	& c	\\
RP4250	&	05 57 42.96	&	-69 02 49.10	&	T	& c	\\
\hline\noalign{\smallskip}
  \end{tabular}\label{table 1}
   \end{center}
   \end{table}
   \begin{table}
\begin{center}
\caption{Table 1 continued}
\begin{tabular}{lllcl}
\hline\hline \noalign{\smallskip}
RP cat. & RA $^{(h~m~s)}$ &  Dec $^{(\circ~_{'}~_{''})}$   & Prob. & Comments  \\
     No.   &  J2000   &  J2000        &          &       \\
  \hline\noalign{\smallskip}
RP4284	&	05 58 38.77	&	-66 56 57.20	&	L	& c,f	\\
RP4285	&	05 58 39.68	&	-65 05 15.10	&	T	& e	 \\
RP4305	&	06 01 42.16	&	-71 08 09.50	&	P	& c	\\
RP4319	&	04 48 17.85	&	-68 44 32.47	&	L	& c, ext?	\\
RP4489	&	06 15 42.71	&	-70 53 32.38	&	L	& e, near star	\\
RP4673	&	05 29 42.97	&	-72 54 56.13	&	T	& c	\\
RP4687	&	05 34 38.74	&	-73 27 19.93	&	L	& e	\\
\hline\noalign{\smallskip}
  \end{tabular}
   \end{center}
  \end{table}


\section{Conclusion}
\label{section5}
Using the MCELS \OIII5007\AA/H$\alpha$/\SII6716-6731\AA~and SuperCOSMOS H$\alpha$/SR image maps we have been able to uncover an extra 63 spectroscopically confirmed PNe in the outer regions of the LMC not covered by our original deep H$\alpha$ stack in the central 25deg$^{2}$  (Reid \& Parker 2006a). Evolutionary theory predicts that there should be approximately 900 PNe in the LMC (Reid 2012). These new discoveries add an additional 9\% to the known LMC PNe population overall but a considerably larger fraction (40\%) of those known in the outer regions.

We have gathered medium resolution spectra and multi-wavelength imaging for each source and put the resulting data through important tests that give us confidence in our classifications. These PNe represent some of the faintest and most evolved PNe in the outer LMC. They have been included in an LMC PN central star temperature histogram and a temperature versus expansion velocity plot shown in Reid (2013).

Such PNe provide valuable data to build up the faint end of the PN luminosity function, to understand the evolutionary time-scales for PNe and in providing additional objects as kinematic and abundance tracers across the LMC. Optical diagnostic plots show that all the newly discovered PNe conform to the ratios found for other LMC PNe. Our paper on multi-wavelength comparisons (Reid \& Parker 2013) is an adjunct to this work and extends to include all PNe in the LMC.

Evolutionary theory predicts that there should be approximately 900 PNe in the LMC (Reid 2012) and we now have 715 LMC PNe recorded in our data base, bringing us very close to evolutionary predictions.

\label{section 9}

\section*{Acknowledgments}

The authors wish to thank the AAO board for observing time on the
AAT and UKST. WR thanks Macquarie University, Sydney, for a research fellowship accompanied by travel grants.

\label{section 10}

\section{Appendix}
Left to right: Column 1: Merged UKST H$\alpha$ (blue) and Short Red (red) discovery images. Column 2: MCELS H$\alpha$ (red), \SII~(blue), and \OIII~(green) discovery images. Column 3: Blue end spectroscopic data using the 580V grating on both AAOmega and 6dF (where indicated). Column 4: Red end spectroscopic data using the 2000R grating on AAOmega and 1000R grating on 6dF (where indicated). The scaling of UKST and MCELS images is approximate due to a difference in viewing angles and pixel sizes.

 \begin{figure*}
\begin{center}
  \includegraphics[width=0.98\textwidth]{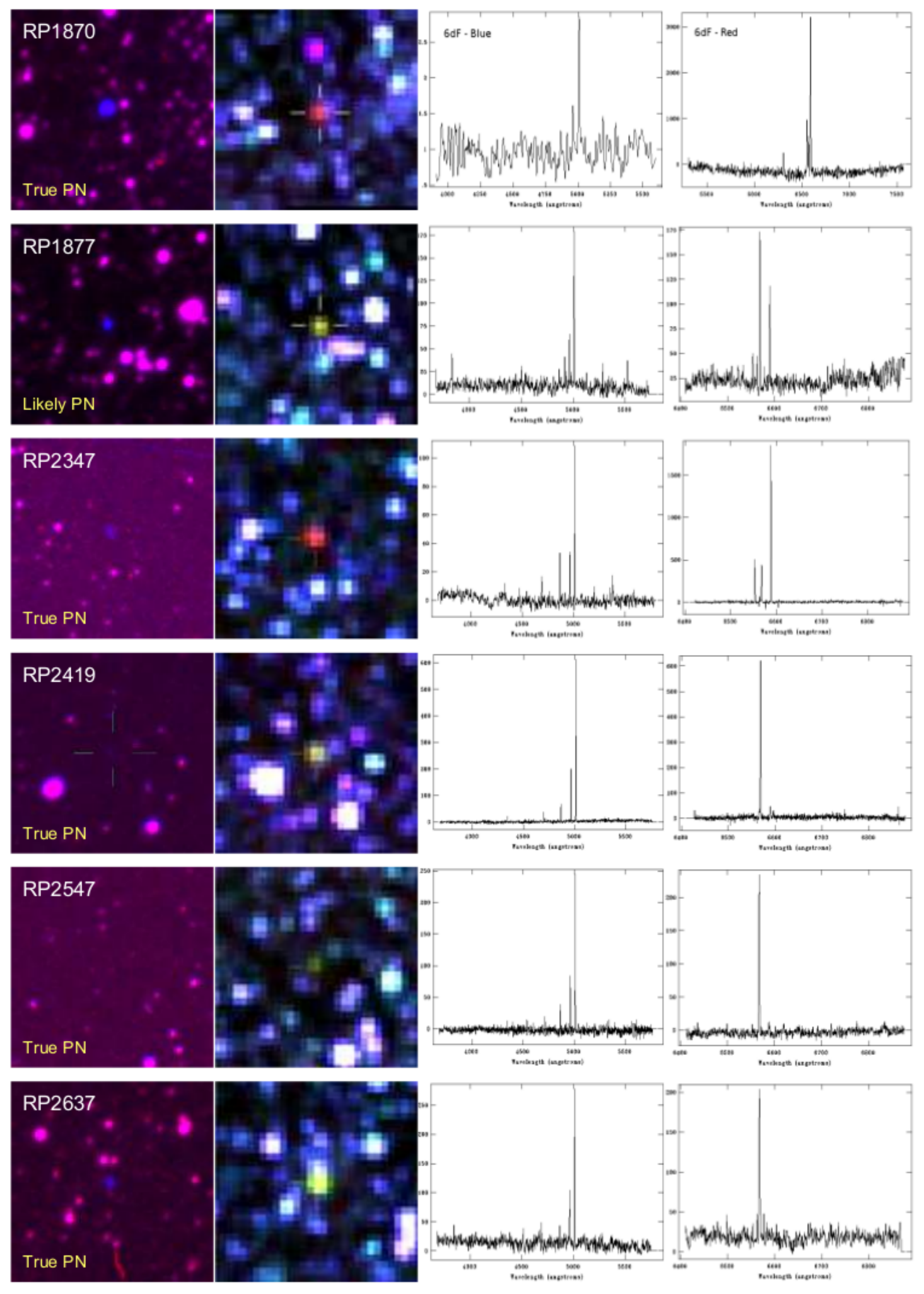}\\
   \label{Figure12}
  \end{center}
  \end{figure*}

\begin{figure*}
\begin{center}
  \includegraphics[width=0.98\textwidth]{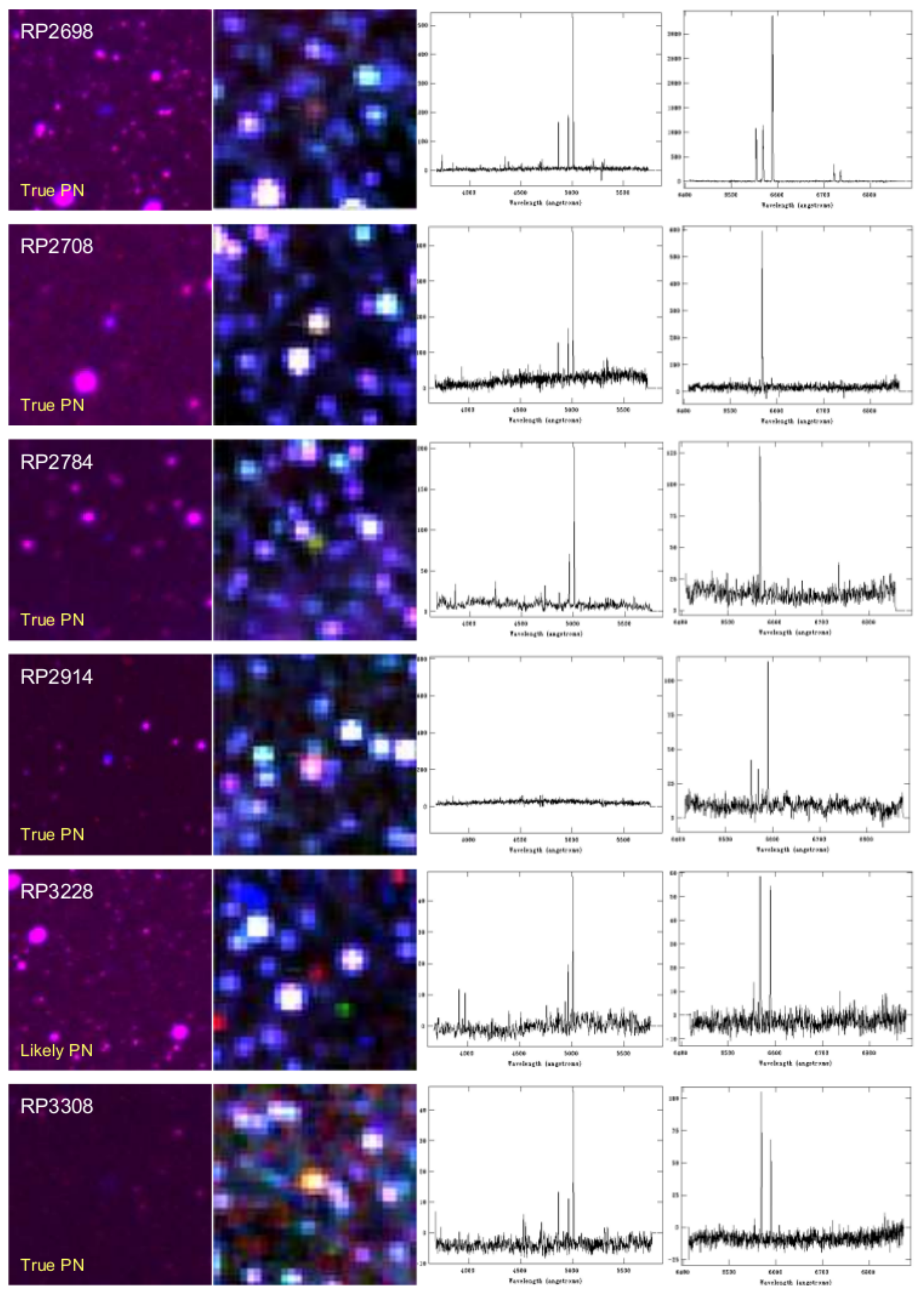}\\
  \label{Figure13}
 \end{center}
  \end{figure*}

  \begin{figure*}
\begin{center}
  \includegraphics[width=0.98\textwidth]{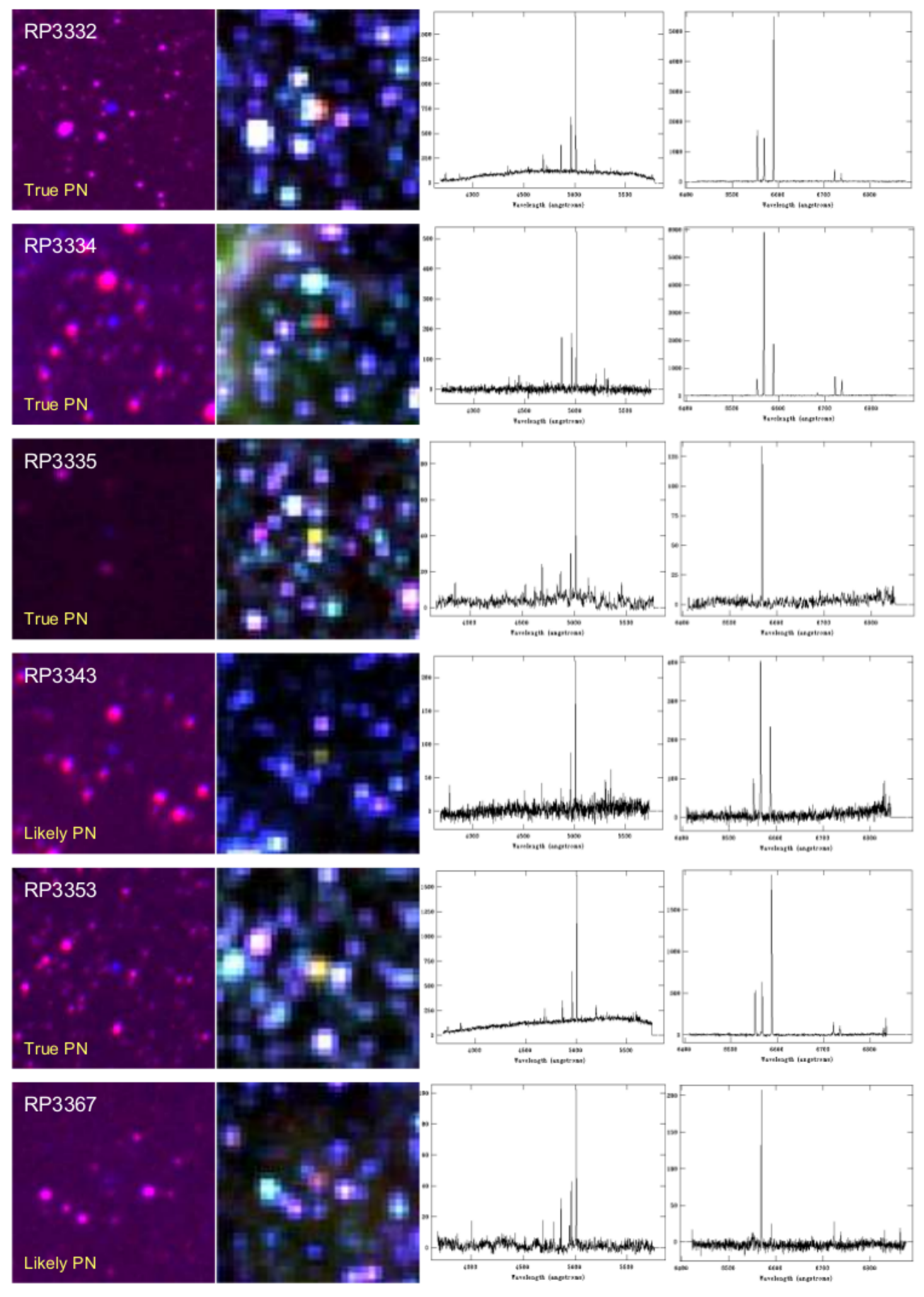}\\
  \label{Figure14}
  \end{center}
  \end{figure*}

  \begin{figure*}
\begin{center}
  \includegraphics[width=0.98\textwidth]{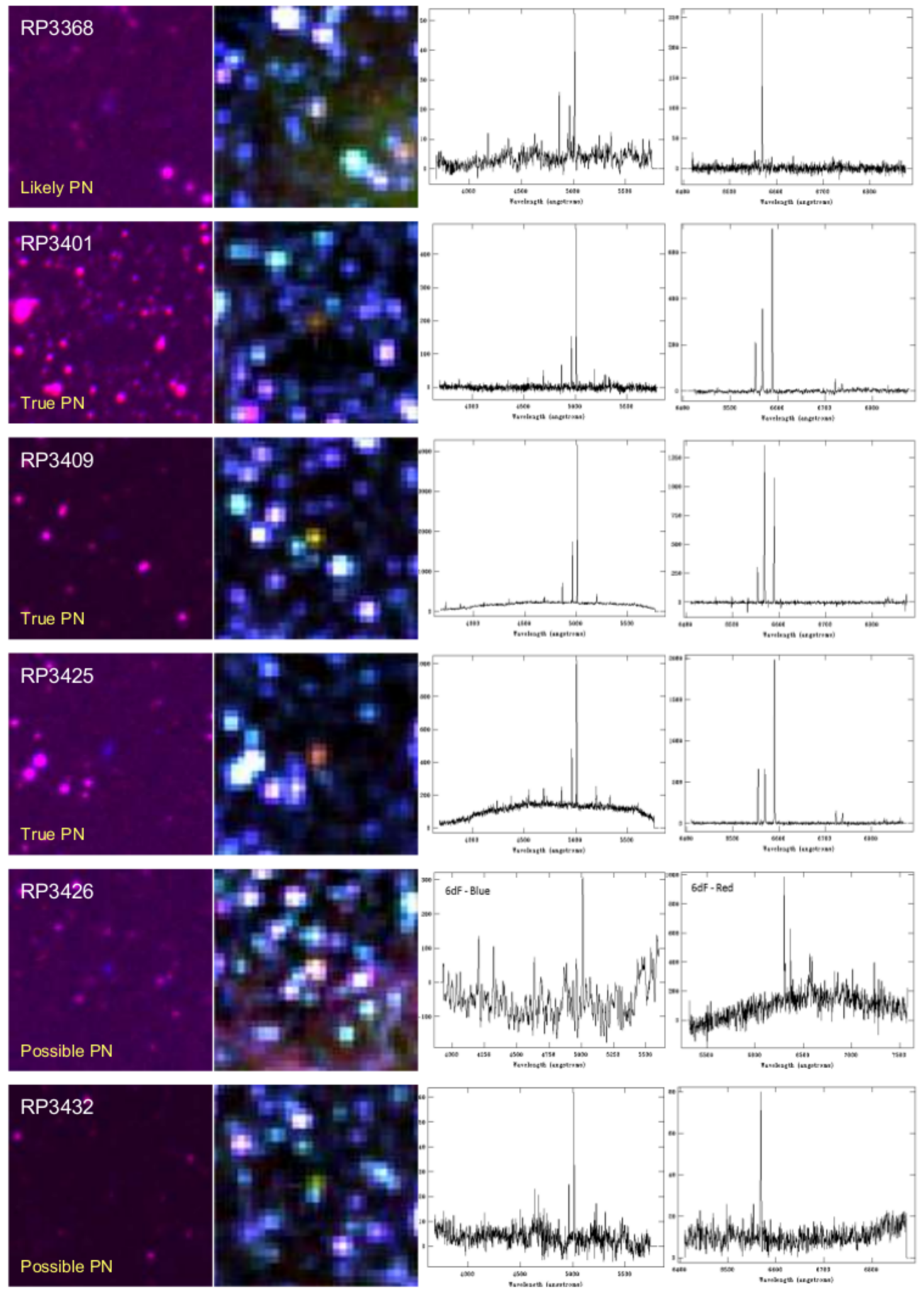}\\
  \label{Figure15}
  \end{center}
  \end{figure*}

 \begin{figure*}
\begin{center}
  \includegraphics[width=0.98\textwidth]{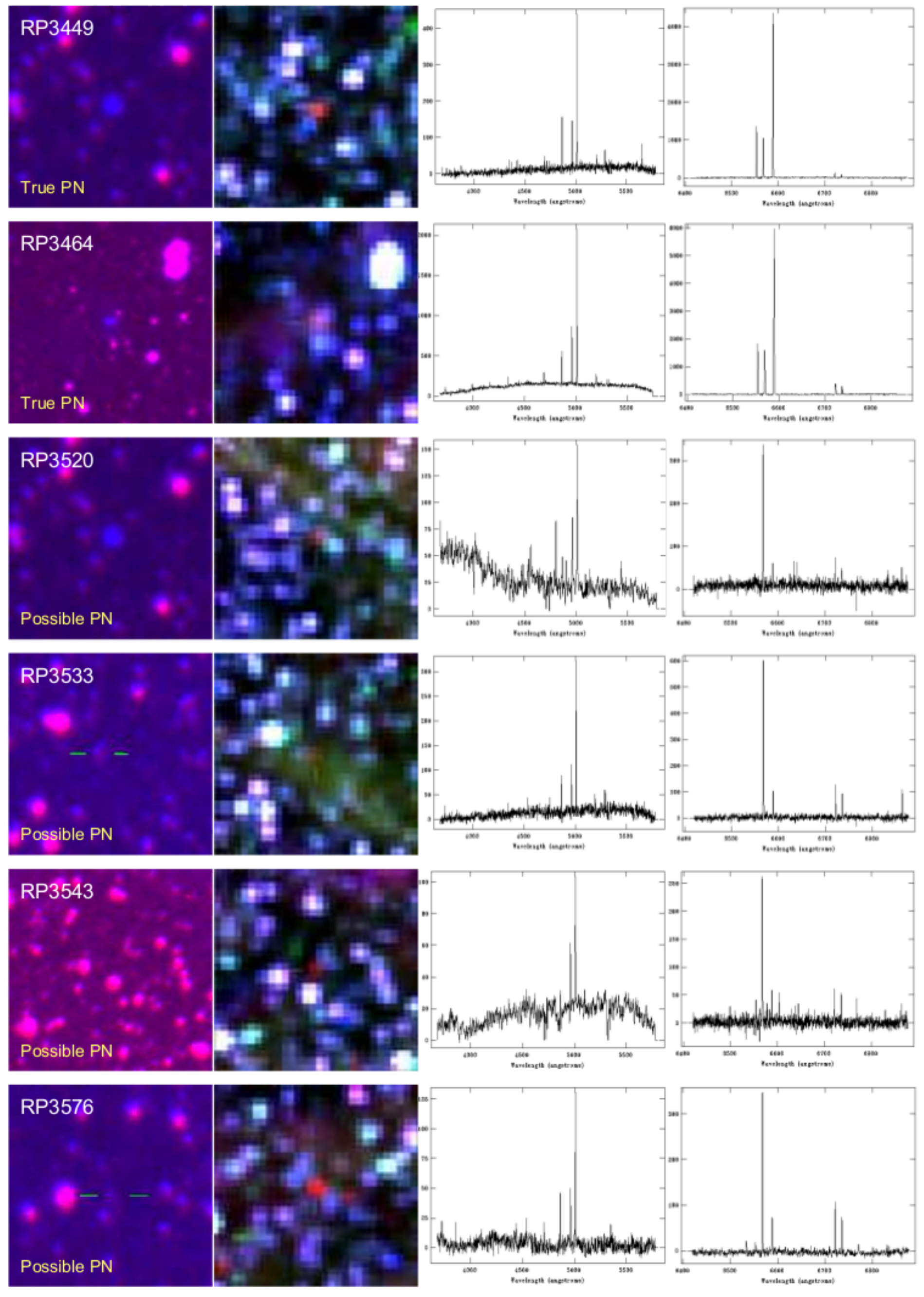}\\
  \label{Figure16}
  \end{center}
  \end{figure*}

  \begin{figure*}
\begin{center}
  \includegraphics[width=0.98\textwidth]{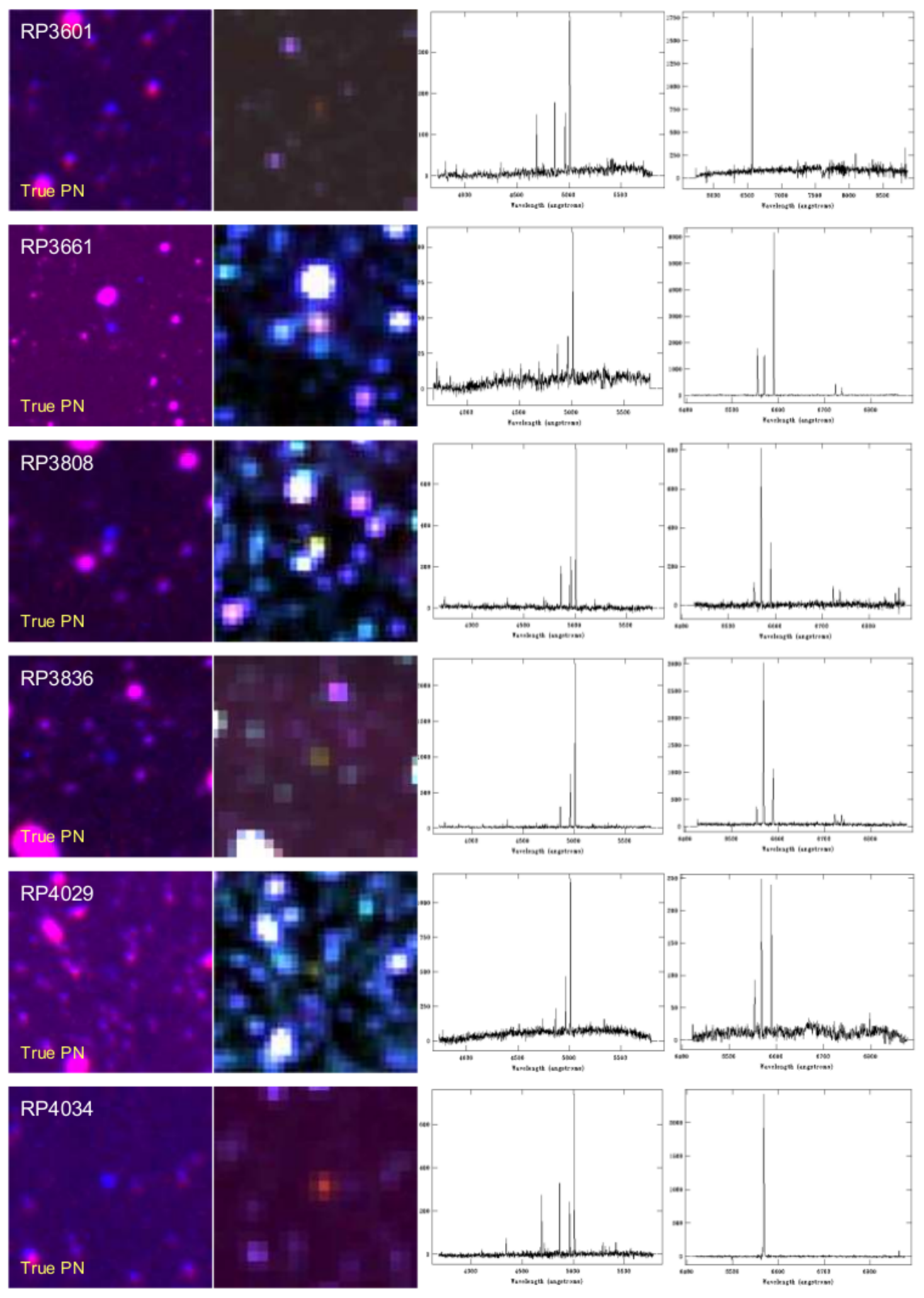}\\
  \label{Figure17}
  \end{center}
  \end{figure*}

  \begin{figure*}
\begin{center}
  \includegraphics[width=0.98\textwidth]{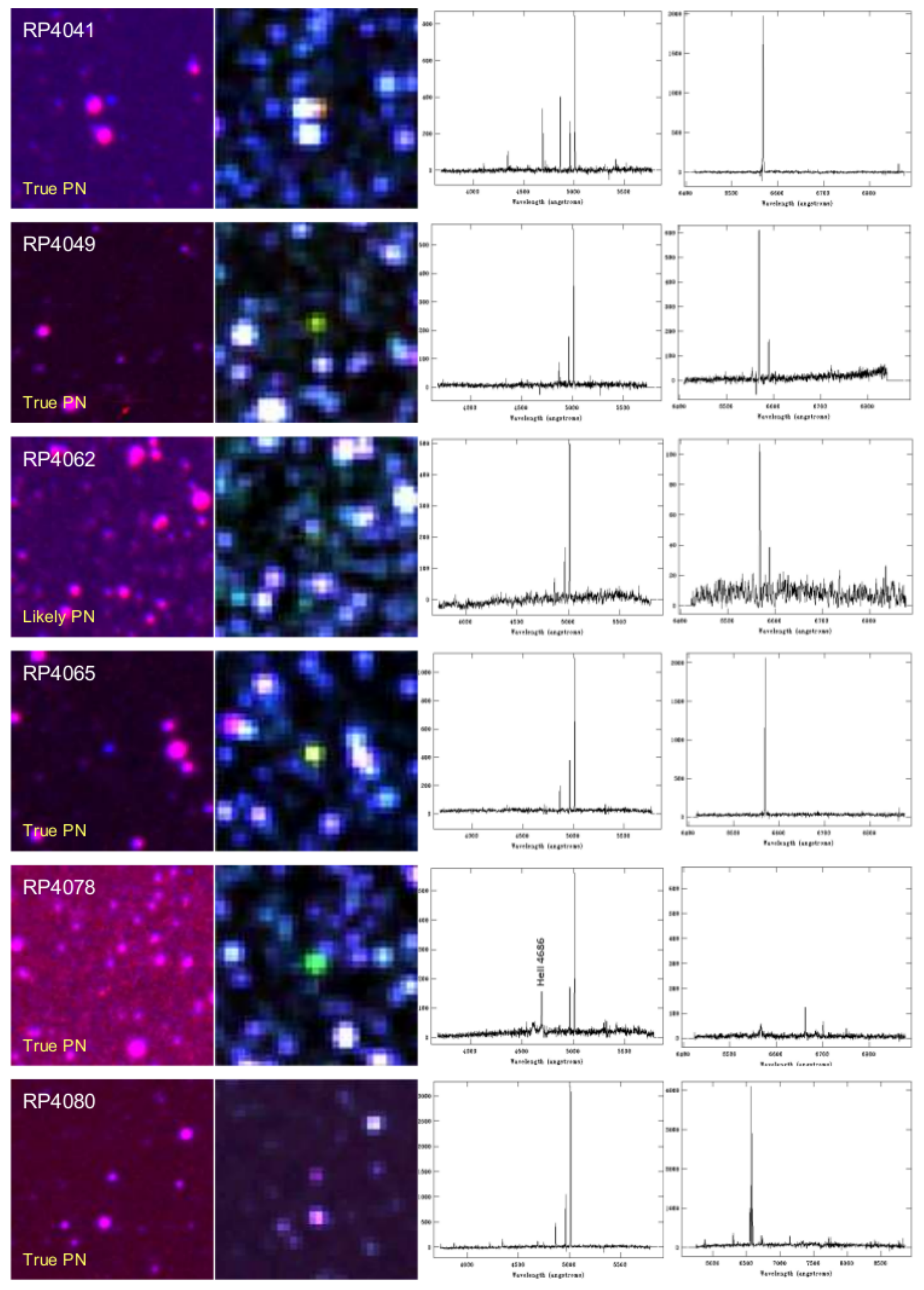}\\
  \label{Figure18}
  \end{center}
  \end{figure*}

  \begin{figure*}
\begin{center}
  \includegraphics[width=0.98\textwidth]{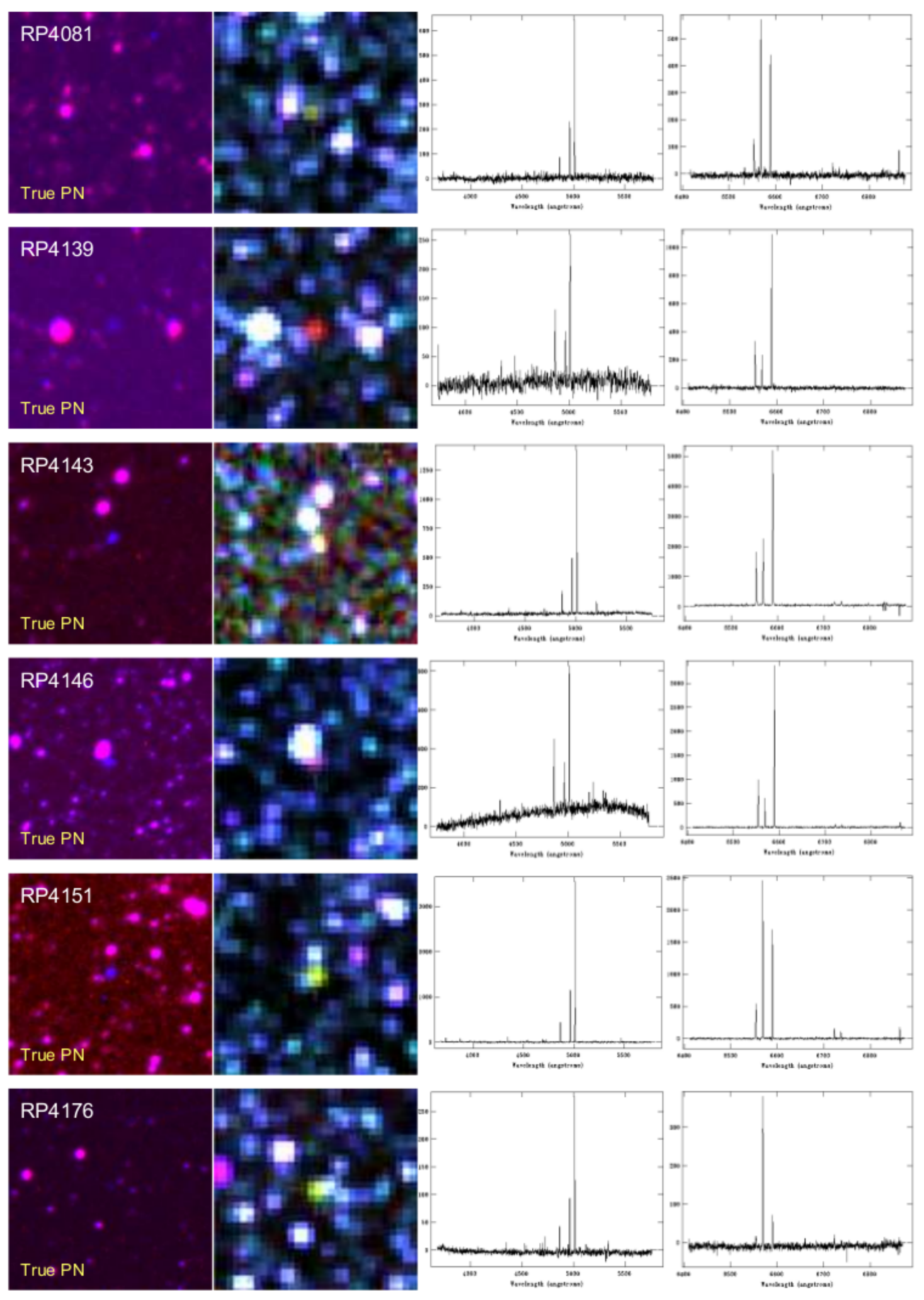}\\
  \label{Figure19}
  \end{center}
  \end{figure*}

  \begin{figure*}
\begin{center}
  \includegraphics[width=0.98\textwidth]{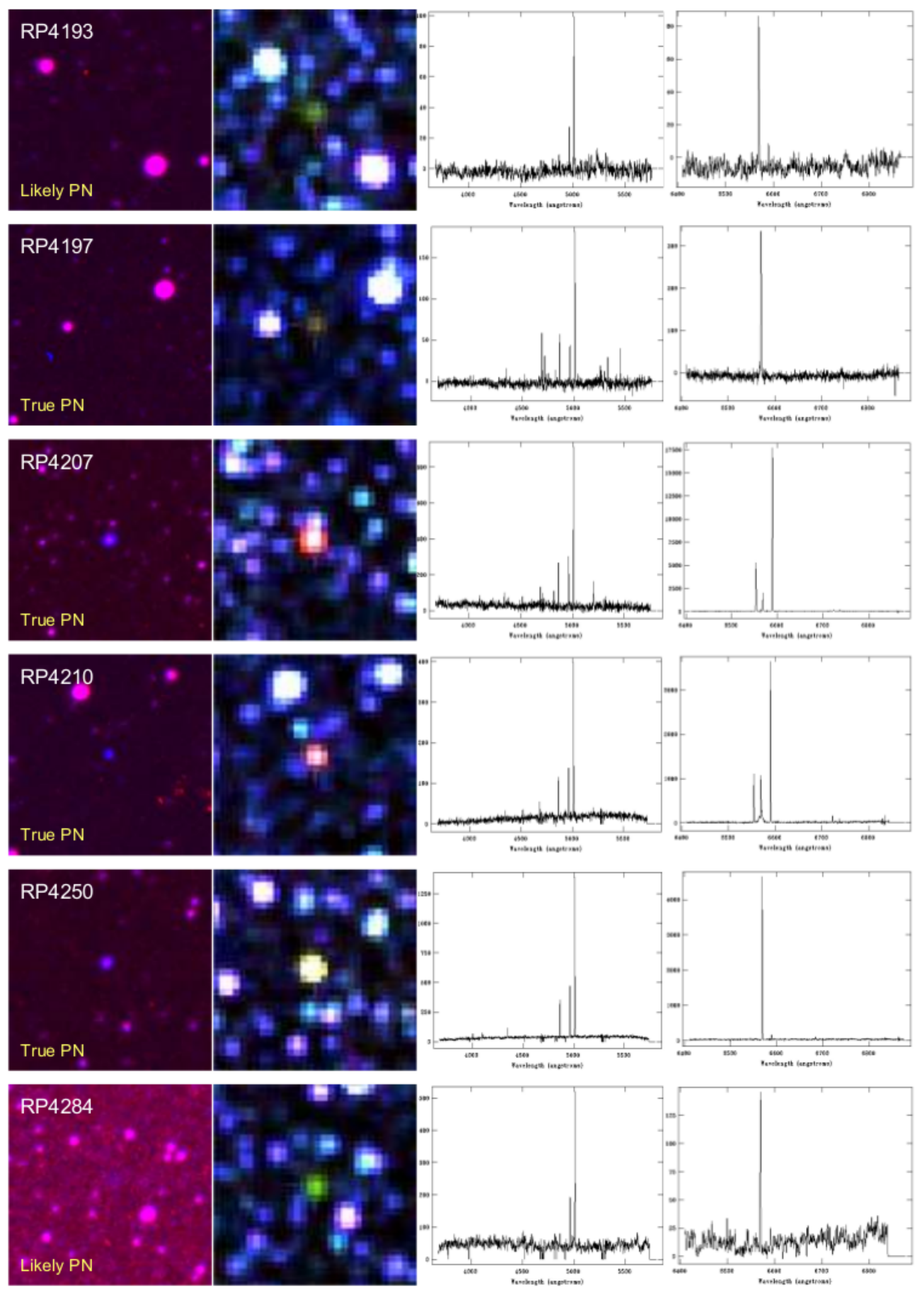}\\
  \label{Figure20}
  \end{center}
  \end{figure*}

  \begin{figure*}
\begin{center}
  \includegraphics[width=0.98\textwidth]{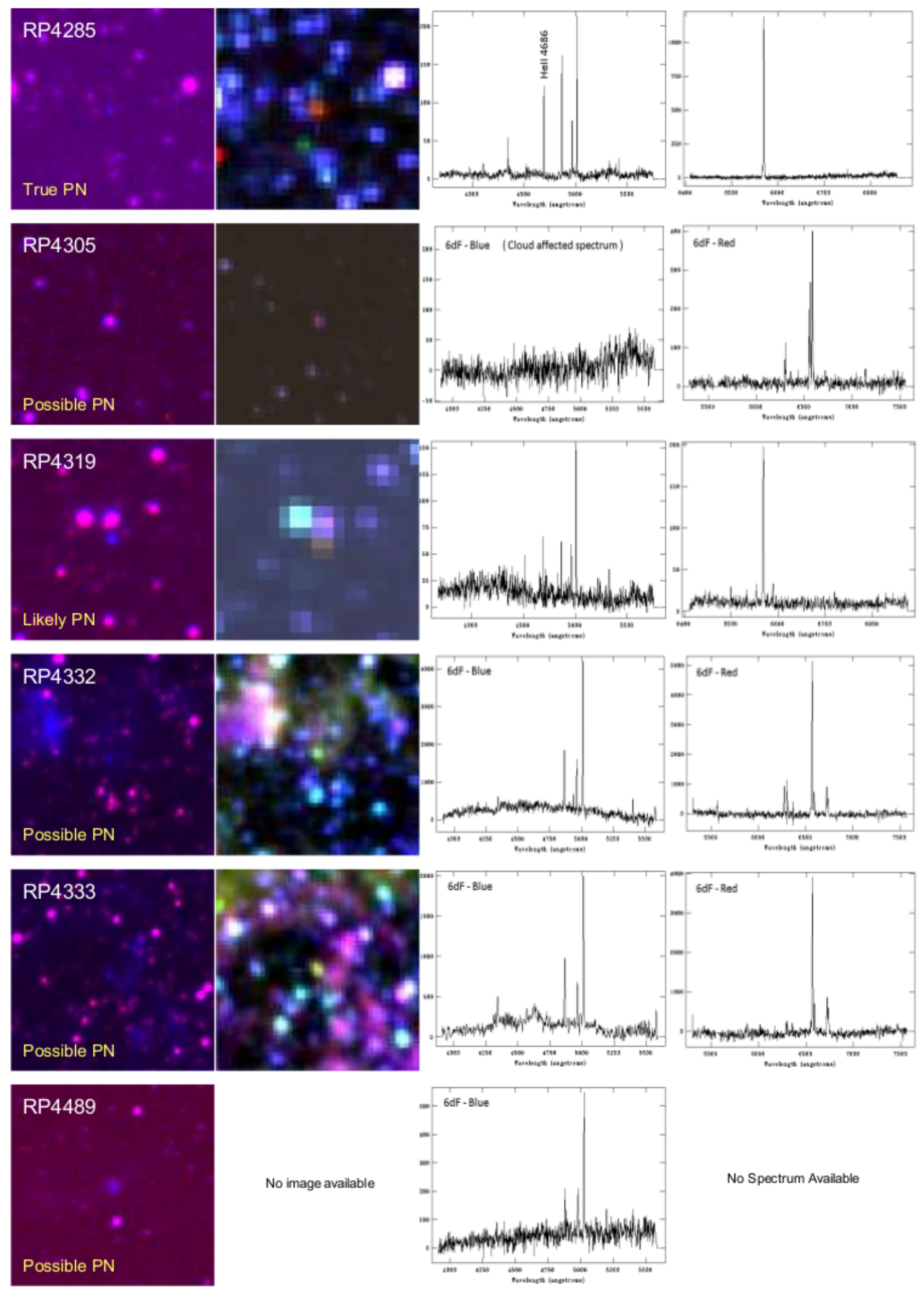}\\
  \label{Figure21}
  \end{center}
  \end{figure*}


\bsp

\label{lastpage}

\end{document}